\documentclass[twoside]{article}
\makeatletter\if@twocolumn\PassOptionsToPackage{switch}{lineno}\else\fi\makeatother

\usepackage{tabulary,graphicx,times,caption,fancyhdr,amsfonts,amssymb,amsbsy,latexsym,amsmath}
\usepackage{braket}
\usepackage{breakurl}
\usepackage{graphicx}
\usepackage[rightcaption]{sidecap}
\usepackage{float}
\usepackage{amssymb}
\usepackage{qcircuit}
\usepackage{tabularx}
\usepackage{subcaption}
\usepackage{cite}

\usepackage[utf8]{inputenc}

\usepackage{url,multirow,morefloats,floatflt,cancel,tfrupee}
\makeatletter

\AtBeginDocument{\@ifpackageloaded{textcomp}{}{\usepackage{textcomp}}}
\makeatother
\usepackage{colortbl}
\usepackage{xcolor}
\usepackage{pifont}
\usepackage[nointegrals]{wasysym}
\makeatletter

\def\mcWidth#1{\csname TY@F#1\endcsname+\tabcolsep}

\def\cAlignHack{\rightskip\@flushglue\leftskip\@flushglue\parindent\z@\parfillskip\z@skip}
\def\rAlignHack{\rightskip\z@skip\leftskip\@flushglue \parindent\z@\parfillskip\z@skip}

\@ifundefined{etal}{}{}

\usepackage{ifxetex}
\ifxetex\else\if@twocolumn\@ifpackageloaded{stfloats}{}{\usepackage{dblfloatfix}}\fi\fi

\AtBeginDocument{
\expandafter\ifx\csname eqalign\endcsname\relax
\def\eqalign#1{\null\vcenter{\def\\{\cr}\openup\jot\m@th
  \ialign{\strut$\displaystyle{##}$\hfil&$\displaystyle{{}##}$\hfil
      \crcr#1\crcr}}\,}
\fi
}

\AtBeginDocument{%
  \@ifpackageloaded{endfloat}%
   {\renewcommand\efloat@iwrite[1]{\immediate\expandafter\protected@write\csname efloat@post#1\endcsname{}}}{\newif\ifefloat@tables}%
}%

\def\BreakURLText#1{\@tfor\brk@tempa:=#1\do{\brk@tempa\hskip0pt}}
\let\lt=<
\let\gt=>
\def\processVert{\ifmmode|\else\textbar\fi}

\@ifundefined{subparagraph}{
\def\subparagraph{\@startsection{paragraph}{5}{2\parindent}{0ex plus 0.1ex minus 0.1ex}%
{0ex}{\normalfont\small\itshape}}%
}{}

\newcommand\role[1]{\unskip}
\newcommand\aucollab[1]{\unskip}
  
\@ifundefined{tsGraphicsScaleX}{\gdef\tsGraphicsScaleX{1}}{}
\@ifundefined{tsGraphicsScaleY}{\gdef\tsGraphicsScaleY{.9}}{}
\def\checkGraphicsWidth{\ifdim\Gin@nat@width>\linewidth
	\tsGraphicsScaleX\linewidth\else\Gin@nat@width\fi}

\def\checkGraphicsHeight{\ifdim\Gin@nat@height>.9\textheight
	\tsGraphicsScaleY\textheight\else\Gin@nat@height\fi}

\def\fixFloatSize#1{}
\let\ts@includegraphics\includegraphics

\def\inlinegraphic[#1]#2{{\edef\@tempa{#1}\edef\baseline@shift{\ifx\@tempa\@empty0\else#1\fi}\edef\tempZ{\the\numexpr(\numexpr(\baseline@shift*\f@size/100))}\protect\raisebox{\tempZ pt}{\ts@includegraphics{#2}}}}

\AtBeginDocument{\def\includegraphics{\@ifnextchar[{\ts@includegraphics}{\ts@includegraphics[width=\checkGraphicsWidth,height=\checkGraphicsHeight,keepaspectratio]}}}

\DeclareMathAlphabet{\mathpzc}{OT1}{pzc}{m}{it}

\def\URL#1#2{\@ifundefined{href}{#2}{\href{#1}{#2}}}

\def\UrlOrds{\do\*\do\-\do\~\do\'\do\"\do\-}%
\g@addto@macro{\UrlBreaks}{\UrlOrds}

\edef\fntEncoding{\f@encoding}

\makeatother

\newif\ifmultipleabstract\multipleabstractfalse%
\usepackage[paperheight=11in,paperwidth=8.3in,margin=2.5cm,headsep=.7cm,top=2.5cm]{geometry}
\usepackage[T1]{fontenc}

\renewenvironment{abstract}
{\vspace*{-1pc}\trivlist\item[]\leftskip\hindawiIndent\par\vskip4pt\noindent\textbf{\abstractname}\mbox{\null}\\}{\par\noindent\endtrivlist}

 \date{} \emergencystretch 8pt

\makeatletter\def\hindawiIndent{0pc}
\def\author#1{\gdef\@author{\hskip-\dimexpr(\tabcolsep)\hskip\hindawiIndent\parbox{\dimexpr\textwidth-\hindawiIndent}{\raggedright\bfseries#1}}}

\def\title#1{\gdef\@title{\vspace*{-30pt}\raggedright\textbf{ \journaltitle}~\\\raggedright\bfseries\ifx\@articleType\@empty\vspace*{20pt}\else\vspace*{20pt}\@articleType\vspace*{20pt}\\\fi#1}}
\let\@articleType\@empty \def\articletype#1{\gdef\@articleType{{\normalfont\itshape#1}}}

\let\@runningHead\@empty \def\RunningHead#1{\gdef\@runningHead{{\normalfont #1}}}

\usepackage{fancyhdr}
\fancypagestyle{headings}{\fancyhf{}\fancyhead[R]{\thepage}\fancyhead[R]{Simulation of non-radiative energy transfer in photosynthetic systems using a quantum computer}\fancyfoot[C]{\thepage}}
\fancypagestyle{plain}{\fancyhf{}\fancyhead[R]{Simulation of non-radiative energy transfer in photosynthetic systems using a quantum computer}\fancyfoot[C]{\thepage}}

\makeatother

\usepackage[numbers,sort&compress]{natbib}

\def\journaltitle{Simulation of non-radiative energy transfer in photosynthetic systems using a quantum computer}
\begin{document}

\title{\mbox{}}
\author{Jos{\'e} Diogo Guimar{\~{a}}es\textsuperscript{1,2},
            Carlos Tavares\textsuperscript{1,3}, Luís Soares Barbosa\textsuperscript{1,3,5} and Mikhail I. Vasilevskiy\textsuperscript{2,4,5} ~\\[-3pt]\normalsize\normalfont 
~\\\textsuperscript{1}{Department of Informatics\unskip, University of Minho, Campus de Gualtar, Braga, Portugal}
~\\\textsuperscript{2}{Department of Physics\unskip, University of Minho, Campus de Gualtar, Braga, Portugal}
~\\\textsuperscript{3}{High-Assurance Software Laboratory\unskip, INESC TEC\unskip, Departament of Informatics, University of Minho, Campus de Gualtar\unskip, Braga\unskip, Portugal}
~\\\textsuperscript{4}{Centro de F\'{\i}sica, Universidade do Minho, Campus de Gualtar, Braga, Portugal}
~\\\textsuperscript{5}{International Iberian Nanotechnology Laboratory, Braga, Portugal}}

\maketitle 

\begin{abstract}

\noindent   
Photosynthesis is an important and complex physical process in nature, whose comprehensive understanding would have many relevant industrial applications, for instance in the field of energy production. In this paper we propose a quantum algorithm for the simulation of the excitonic transport of energy, occurring in the first stage of the process of photosynthesis. The algorithm takes in account the quantum and environmental effects (\emph {pure dephasing}), influencing the quantum transport. We performed quantum simulations of such phenomena, for a proof of concept scenario, in an actual quantum computer the IBM Q, of 5 qubits. We validate the results with the Haken-Str\"{o}bl model and discuss the influence of environmental parameters on the efficiency of the energy transport.

\end{abstract}

\section {Introduction}

Photosynthesis is a vital and pervasive complex physical process in nature, where the radiation of the Sun is captured by certain \emph {living beings}, such as plants and bacteria, and transformed into the necessary \emph {carbohydrates} needed for their survival \cite{mohseni2014quantum, lambert2013quantum}. From the physics and chemistry perspective, it is a complex process occurring through several stages with several kinds of physical phenomena involved, namely, the light absorption, energy transport, charge separation, photophosphorylation and carbon dioxide fixation \cite {fleming1994primary}. The understanding of such phenomena has greatly progressed in the the past 40 years with the physical characterization of the structure of many photosynthetic complexes \cite {deisenhofer1989photosynthetic, schubert1997photosystem, cheng2006coherence}. The comprehension of such processes would allow for many potential huge-impact industrial \emph {breakthroughs} in the field of energy, from the great efficiency improvement in energy capture of solar panels \cite {lewis2016research} to the construction of artificial light-harvesting devices and solar fuels \cite {xu2018artificial, gust2001mimicking, gust2009solar, romero2017quantum}.

\begin{figure}[H]
        \centering
        \includegraphics[width=0.8\linewidth]{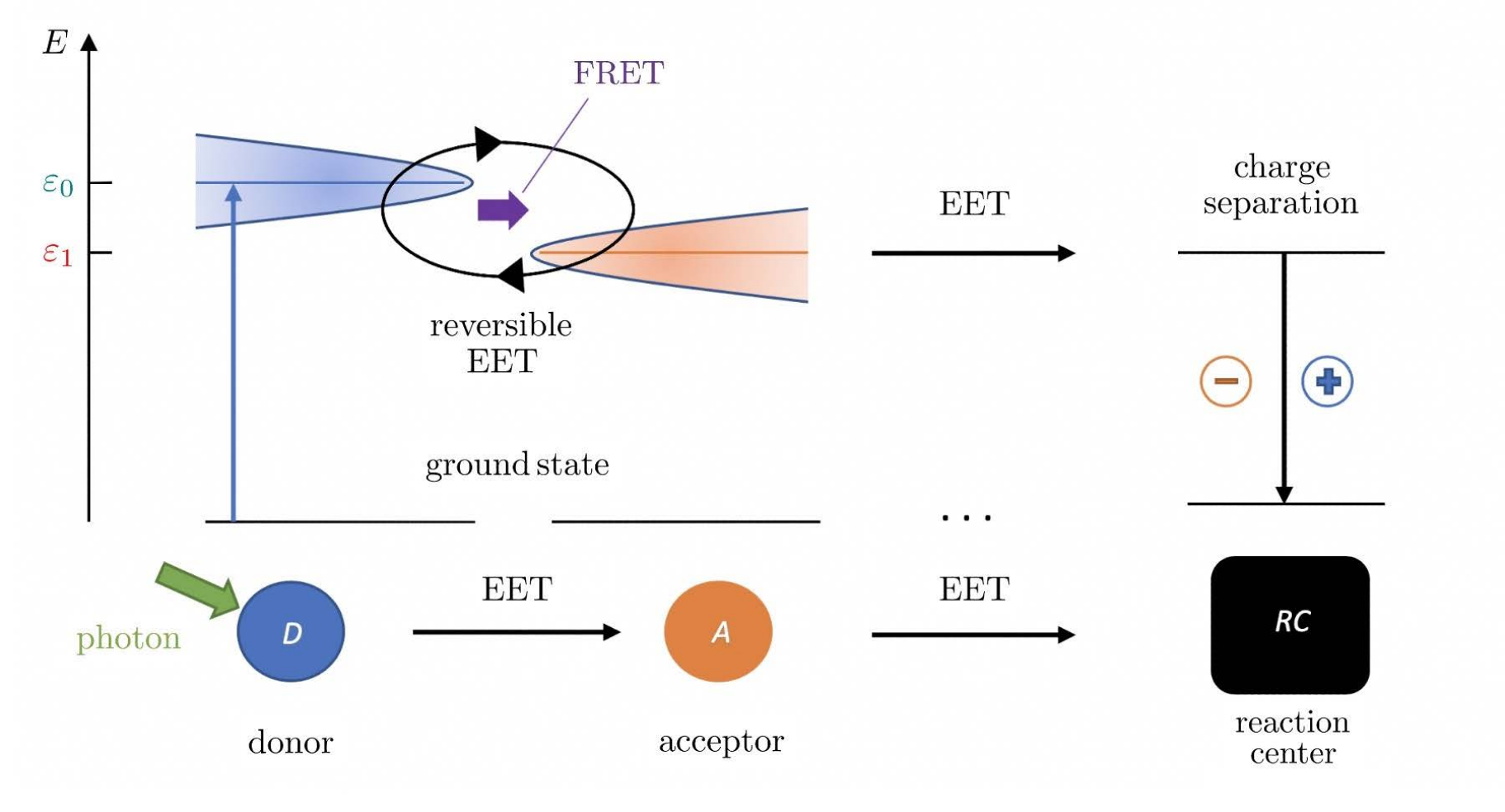}
        \caption{Schematics of the energy transfer process from light-harvesting antenna (the donor) through a chain of acceptor molecules to the reaction center. The excited states of the participating molecules, denoted $\epsilon _m$, are broadened and it allows for resonance energy transfer via irreversible F\"orster-type resonant process of exciton transfer from donor to acceptor even if $\epsilon _m \neq \epsilon _{m+1}$, which is denoted by the thick arrow labelled FRET. However, if the coupling between the donor and the acceptor molecules is strong enough, the process becomes reversible and the exciton can go to and through many times before it is transferred; this situation is labeled by "reversible EET" and it does not require matching of the energy levels $\epsilon _m$ and $\epsilon _{m+1}$.}
        \label {fig:scheme}
    \end{figure}

The photosynthesis begins by the absorption of a photon. It occurs via excitation of a pigment molecule, which acts as a \emph {light-harvesting antenna} connected to the rest of the photosynthetic apparatus by protein molecules. Photosynthetic pigment-protein complexes transfer the absorbed sunlight energy, in the form of molecular electronic excitation, to the reaction center, where charge separation initiates a series of biochemical processes \cite {mohseni2014quantum}.
This work is focused on the first stage of photosynthesis, more precisely, on the transport of the absorbed radiation energy from the antenna to the reaction centre, which proceeds in the form of the so-called Excitonic Energy Transfer (EET), as schematically shown in Fig.\ref {fig:scheme}.  

This transport is known to be very efficient in photosynthesis, as is the whole process, with the overall quantum efficiency of initiation of charge separation per absorbed photon up to 95\% \cite {mohseni2014quantum}. The absorbed photon creates an \emph {exciton} on the antenna molecule, which can eventually transfer it to other molecules. In this context, it is called donor, while the others are called acceptors and the EET process can be described by the following reaction equation:
\begin {equation}
D^\star +A \rightarrow D + A^\star \, .
\label {eq:EEt}
\end {equation}

The physics of the mechanisms behind Eq. (\ref {eq:EEt}) will be discussed in the following section. Here we just notice that EET is a complex process that can be irreversible (i.e. unidirectional) or reversible, i.e. coherent over some period of time as evidenced by experimental observations of long-lived oscillatory features in the dynamical response of several photosynthetic systems \cite{engel2007evidence,lee2007coherence,hildner2013quantum}. 
Moreover, it is strongly influenced by the environment. The donor-acceptor pair is not isolated from the rest of the world and is an example of so called \emph {open systems}  \cite{rebentrost2009environment}. It must be treated as a subsystem of a larger system including a thermal bath. The properties of the latter are crucial because it introduces relaxation and dephasing into the system directly involved in the EET and, therefore, influences the efficiency of the energy transport.

Open quantum systems cannot be described by a wave function because one does not have enough information to specify it, only a (less detailed) description in terms of a density matrix is possible, which represents a statistical mixture of states, or a mixed state (see Supplementary information A.1). The dynamics of such a system can be determined by solving an equation of motion for the density matrix. Such equations of motion are called \emph {quantum master equations}. Finding exact solutions to the master equations is extremely difficult but there is a wide range of theoretical approaches and techniques available to make their mathematical simplification and numerical simulations.
These approaches can be divided into several groups according to the regime under study, characterized by the {coupling strength} between the \emph {bath} and the \emph {system}, and the existence of \emph {memory} effects in the bath (i.e. whether the system can be considered as Markovian or not). Broadly speaking, for the \emph {weak-coupling} \emph {Markovian} regimes,  perturbative approaches are applicable, such as the Bloch-Redfield and Lindblad master equations \cite {mohseni2014quantum,jeske2015bloch}, which can be extended to {medium} {coupling strengths} and \emph {non-Markovian} regimes by including  higher-order system-bath interaction terms \cite {reichman1996relaxation, jang2002fourth}. For the latter regime, there are also \emph {non-perturbative} techniques based on the use of \emph {path integrals} to \emph {dissipative} systems \cite { feynman1963theory}, which can be used to create sets of solvable systems of hierarchical equations, the so-called hierarchical equations of motion (HEOM) \cite {tanimura2006stochastic, tanimura1989time, ishizaki2008nonperturbative}.
Open quantum systems that do not have the Markovian property, for example, because of a too small size of the bath effectively coupled to the system, which keeps memory of the past, are much more difficult for theoretical description because the dynamics equation is non-local in time.

However, even within the Markov approximation, the calculations quickly become computationally intractable for realistic photosynthetic systems and environment models. The computational cost of simulating a photosynthetic complex consisting of $N$ molecules with a theoretical tool such as the HEOM grows exponentially with $N$. A possible computational solution has been arising to bypass this type of problems is the use of quantum simulation, where it is expected to obtain large performance increases in terms of \emph {space}, as the number of qubits' growth is just polynomial, and in terms of \emph {time}, where an exponential gain is expected.

The use of quantum mechanics to make calculations about quantum mechanics, promising great computational advantages was firstly proposed by R. Feynman \cite {feynman1982simulating, lloyd1996universal}. The field of quantum simulation is under a fast paced and intense development, finding already application across all fields of physics and using many different \emph {physical implementations} \cite {georgescu2014quantum}. Closer to the present work, there are works on quantum transport \cite {maier2019environment, schonleber2015quantum} and on the quantum simulation of dissipative systems \cite {di2015quantum}. Particularly on the quantum simulation of photosynthesis, we would like to highlight Refs. \cite {mostame2012quantum, potovcnik2018studying,tao2018quantum}, using {superconducting qubits}, and \cite{wang2018efficient} employing a Nuclear Magnetic Resonance (NMR) simulator \cite {zhen2016optimal}. The latter is of particular relevance to the work carried out here as it was dedicated to the quantum  simulation of the energy transport with environmental actions, where the environment effect is simulated naturally by an appropriate filtering of environmental noise \cite {soare2014experimental}, within the NMR system.  In this case, the implementation is specific for the EET (i.e. non-universal), and the  model Hamiltonian was extracted from spectroscopic data for a photosynthetic system \cite {ai2013clustered}. 
 We are simulating the same Hamiltonian as in Ref. \cite{wang2018efficient} and starting from the same assumptions, however, the simulation algorithm is completely different since we conceived a \emph {digital} quantum simulation designed to run in a universal quantum computer, the {commercially available} IBM Q of 5 qubits \cite {cross2018ibm}. Our implementation contains a quantum part, aimed at simulating the unitary part of the system's evolution, and a classical part that simulates the \emph {stochastic } interaction with the environment, the latter only being able to mimic \emph {pure dephasing} environmental effects.  

\section {The physics of the energy transport in photosynthesis}

\subsection{F\"orster and Redfield approaches}

The molecules of the light-harvesting complexes usually are not electronically coupled to each other and charge transfer via electron tunneling is improbable.  
Hence, energy transfer can occur between them through electromagnetic interaction, without net charge transport because the whole (neutral) exciton is transferred.
Such processes are known to take place between molecules \cite{Andrews_FRET} or artificial nanostructures such as quantum dots \cite{Santos2008} if appropriate conditions are met, which were first formulated by T. F\"orster \cite{forster1965delocalized}:

(i) The distance between the donor and acceptor molecules must be sufficiently small because the transfer probability decreases quickly with the distance between them ($R$), usually as $R^{-6}$;

(ii) There must be a resonance between the excited states of the donor and acceptor molecules;\footnote {"Resonance" here means that the energy spectra of the two molecules, broadened because of a number of natural reasons, overlap - see Fig.\ref {fig:scheme}} 

(iii) An increase of the refractive index of the surrounding medium decreases the transfer rate.

F\"{o}rster's approach is based on the second-order perturbation theory (the so called "Fermi's Golden Rule"), where the perturbation operator is the electromagnetic interaction between two transient dipoles corresponding to allowed optical transitions in the donor and acceptor molecule, respectively. It originated the term "F\"{o}rster resonance energy transfer" (FRET), which applies to an \emph {irreversible} hopping of an exciton from the donor to the acceptor. The FRET rate (transition probability per unit time) can be expressed by the following relation \cite{mohseni2014quantum}:

\begin{equation}
        k_{F} = \frac{J^{2}}{2\pi \hbar^2}\int_{-\infty}^{+\infty} d\omega L_{D}(\omega) I_{A}(\omega)
        \label {eq:FRET}
\end{equation}

\noindent where $J$ is the coupling constant, $\omega$ is the angular frequency of the electromagnetic field and $L_{D}(\omega)$ and $I_{A}(\omega)$ denote dimensionless lineshape functions of the donor and acceptor molecules, directly related to the energy spectrum of each molecule. The integral is called the spectral overlap between the molecules. The coupling constant, in the dipole-dipole approximation, is given by \cite{Andrews_FRET}:
\begin{equation}
        J=\left \langle \frac 1 {\eta ^2 R^3}[((\mbox {d}_A\cdot \mbox {n})\cdot \mbox {d}_D)-3(\mbox {d}_A\cdot \mbox {n})(\mbox {d}_D\cdot \mbox {n})]\right \rangle \;,
        \label {eq:J}
\end{equation}
where $\eta$ is the refractive index of the medium, $\mbox {d}_D (\mbox {d}_A)$ is the transient dipole moment of the donor (acceptor) molecule, $\mbox {n}=\mbox {R}/R$, $\mbox {R}$ is the radius vector between the two molecules and the angular brackets stand for angular average over different orientations of the dipoles.  

Even though Eq. (\ref {eq:FRET}) (and the approach itself) is too simplistic to describe all possible situations in EET, defining this characteristic transfer rate allows for the formulation of the following conditions for FRET to occur:

\begin {itemize}
\item If the difference between the energy of the excited state energy of the donor ($\epsilon_{0}$) and acceptor ($\epsilon_{1}$) molecules is small, $\vert \epsilon_{0}-\epsilon_{1}\vert \ll J$ and they are in resonance, the energy transfer between the molecules can occur with a high probability;
\item If $|\epsilon_{0}-\epsilon_{1}| \gg J$ (off-resonance), the exciton is trapped in the donor molecule because it has a very low probability of being transferred; in this case it either stays in the molecule and later the donor molecule will decay to the ground state, dissipating the energy, or transfer the energy to a different acceptor nearby.
\end {itemize}

As pointed out above, the initial idea of F\"{o}rster was that an exciton is irreversibly transferred from a donor to an acceptor. More recently, it has been shown experimentally that quantum \emph {coherent} transport, where energy is transported in the form of {wave-packets}, has a significant role in many important physical effects, including the photosynthesis \cite {panitchayangkoon2010long, chenu2015coherence, collini2010coherently}. The F\"{o}rster theory does not apply in this regime, as it simply ignores coherence. 
Later, in 1957, A. Redfield \cite {redfield1957theory} proposed a transport theory, which applies to the opposite regime of strong coupling between the donor and acceptor \cite{mohseni2014quantum,chenu2015coherence, leegwater1996coherent} (although originally it appeared in the context of NMR spectroscopy). Within this concept, the exciton forms a coherent state based on the whole donor-acceptor pair and oscillates between the two molecules.
This system can be described by the following Hamiltonian:
\begin{equation}
    \hat{H}_{S}=\sum_{m=0}^{1} \epsilon_{m} \Ket{m} \Bra{m} +J\left (\Ket{0}\Bra{1} + \Ket{1}\Bra{0}\right )\;,
    \label{eq:Ham-2}
\end{equation}
where $\Ket{m}$ denotes the exciton on the molecule $m$. The eigenstates of (\ref{eq:Ham-2}) are linear combinations of $\Ket{0}$ and $\Ket{1}$.
Coherent dynamics corresponds to the presence of non-zero \emph {off-diagonal} elements in the density matrix describing the evolution of the quantum system.
Their oscillation (or \emph {quantum beating}) is indicative of coherence \cite{mohseni2014quantum}. 
For the system with Hamiltonian (\ref{eq:Ham-2}), the description in terms of state vectors is perfectly possible but it will not be the case if interactions with environment are taken into account. Therefore, we may introduce the density matrix description at this point.  
The evolution of the off-diagonal elements of the system's density matrix, written in the \emph{energy basis} (where the Hamiltonian is diagonal) and denoted $\rho_{ij}$, is given by:
\begin{equation}
    \rho_{ij}(t) = e^{-{it} \sqrt{(\epsilon_{0}-\epsilon_{1})^{2}+4J^{2}}/\hbar }\rho_{ij}(0)\,,\qquad i \neq j \,.
    \label{eq:rho}
\end{equation}
\noindent (see {Supplementary Information A.2} for the derivation).
These states are perturbed by interactions with the environment (the bath), which destroys their coherence. Mathematically, it is expressed in the form of a master equation, which is known as the Bloch-Redfield equation; its general form can be found e.g. in Ref. \cite{jeske2015bloch}.

This consideration is extendable to a chain of molecules and can be seen as (partially) coherent transport \cite{chenu2015coherence}. The presence of the latter, observable through coherent oscillations of the energy levels of molecules across different sites ({the quantum beating}), was first conjectured in the 30's \cite {perrin1932theorie} and theoretically predicted in more recent works \cite {knox1996electronic, leegwater1996coherent}. It became possible to observe them more recently, thanks to the advances of optical spectroscopy techniques \cite {van2006energy, lee2007coherence, engel2007evidence, hildner2013quantum}, and it was achieved even at room temperature \cite {panitchayangkoon2010long}. In these experiments, it was possible to confirm the substantial impact of such coherent effects on the excitation energy transfer in photosynthetic systems \cite{mohseni2014quantum}.  Moreover, the importance of environmental noise in the quantum transport involving coherence was also discussed more recently \cite {cao2009optimization, rebentrost2009environment} and it is not fully understood yet.

\subsection{Decoherence}

Processes caused by the molecules' environment may destroy coherence and thus influence this type of energy transport \cite {mohseni2008environment, rebentrost2009environment, mohseni2014quantum}, moreover, they can foster it. Indeed, completely coherent oscillations (called Rabi floppings in atomic physics) between different molecular sites do not correspond to an energy flux. Breaking the oscillatory evolution at some moment may help transferring the exciton along the molecular chain.

If interactions exist between a system and its environment, they affect the (pure) states of the system, introducing  {"errors"} and making these states \emph {mixed}. It means the so-called phenomenon of \emph {decoherence}, which, by the way, has been the main obstacle to the success of quantum computation.
Decoherence processes can be divided into three categories: (i) amplitude damping, (ii) dephasing, and (iii) depolarization, which are briefly described below.\cite{breuer2002theory} 

\textbf{Amplitude damping}. Environment interactions with the system may cause a loss of the amplitude of one or more system's states. The spontaneous emission of a photon from the system (i.e. from one of the molecules) to the environment is an example of this kind of process, so that the system returns to its ground state (without exciton) \cite{nielsen2002quantum}. For a two level system (e.g. a qubit), this type of decoherence contracts the Bloch sphere along the $z$ axis (see Supplementary Information A.1).

\textbf{Phase damping or dephasing}. Such interactions conserve the energy of the system, contrary to the amplitude damping. A phase damping channel removes the superposition of the system state, i.e. the off-diagonal terms of the system's density matrix decay over time down to zero. It is a process of removing the coherence of the system, causing a classical probability distribution of states and, therefore, imposing some classical behaviour in a quantum system. A simple way to look at this type of decoherence is also to think of the system interacting with the environment where the relative phases of the system's states become randomized by the environment. This randomness comes from a distribution of energy eigenvalues of the environment. As a result, the evolution of the quantum system's Rabi cycle ceases but the time-average populations of the states may not change and this is the case of the \emph {pure dephasing}. For a two-level system with a pure dephasing interaction, the Bloch sphere contracts in the $x-y$ plane.

\textbf{Depolarization}. This type of decoherence changes system's state, which initially is pure, to a mixed state, with a probability $P$ of another pure state and the probability $(1-P)$ of the initial state of the system. It is equivalent to saying that, for a single qubit, an initial pure state represented on the Bloch sphere has suffered a contraction over all dimensions of the sphere (with the contraction degree that depends on the probability $P$). It can be thought of as a combination of the other two types of decoherence.

The amplitude damping is certainly detrimental for EET since the energy is simply dissipated into the environment.
The action of {dephasing} processes progressively eliminates the coherence (off-diagonal) elements in the system's density matrix, causing the oscillation amplitude to decay (\emph {beating  supression}). It eventually turns the diagonal matrix elements (populations) into (non-correlated) classical probabilities, a process known as thermal relaxation, for which the existence of coherence in a system is time limited.
On the other hand, it has also been shown that dephasing processes can have a \emph {positive} role in the coherent transport of energy \cite{mohseni2014quantum}. First, it yields random fluctuations in the energy spectrum of each molecule, which can bridge the energy gap between the molecules, momentarily turning a non resonant system into a resonant one. Secondly, {dephasing} can also help avoiding the existence of the so called \emph {coherence traps} in a molecular chain, a kind of \emph{deadlocks} in energy transport where the exciton can be confined \cite{mohseni2014quantum}.
Thus, the result of action of a decoherence source on an EET system is not obvious \emph {a priori}. Below we shall consider a simple model of pure dephasing consisting in a telegraph-type classical noise affecting the donor-acceptor pair. 

\section{Materials and Methods}
\label {models}

We aim at exploring the energy transport underlying the photosynthesis, throughout time, under two regimes: (i) in an isolated system and (ii) under an action of the environment causing decoherence. In the "no decoherence" case (i), one can study the evolution of system's state vector, which obeys the equation

\begin {equation}
\ket {\Psi_{t_f}} = e^{-i\hat{H}{(t_f-t_i)}/\hbar} \ket {\Psi_{t_i}}\equiv \hat{U}\ket {\Psi_{t_i}}
\label {eq:evolution}
\end {equation}

\noindent for a time-independent Hamiltonian. Here $t_f$ and $t_i$ are the upper and lower limits of the time interval to study. In order to be able to do the calculation of the system's evolution on a quantum computer, it is necessary to provide a suitable qubit encoding for the possible states and an approximation for the Hamiltonian evolution operator, $\hat{U}$, in terms of quantum gates and circuits (computational Hamiltonian). The time, a \emph {continuous} entity in equation \eqref {eq:evolution}, has to be \emph {discretized} onto a set of intervals, $\Delta t$, where the Hamiltonian of interest can be approximated as constant. The actual computational process is given by the repeated application of the evolution operator on the \emph {prepared} state $\ket {\Psi_i}$, for $s$ times, of the \emph {computational Hamiltonian}, such that $t_i + s\cdot \Delta t = t_f$. The process is finished by the observation of the desired properties, i.e. a set of measurements, in the appropriate basis, on the end state. 

Concerning the particular qubit encoding chosen, a chain of $N=2^q$  molecules is encoded by a set of $q$ qubits, where $\Ket{m}$ corresponds to the excitation (exciton) on the $m$-th molecule, e.g. for a two-molecule chain, state $\ket {0}$ represents the exciton on the first molecule and $\ket {1}$ on the second one, and a possible successful transport of energy would correspond to the transition of the state $\ket {0}$ to the state $\ket {1}$. We denote this as the \emph {site basis}. The computational Hamiltonians under this encoding for the cases under study are discussed in the following sections. From now on, we shall set $\hbar =1$. Also, it is convenient to measure the energies/frequencies in $\text {cm}^{-1}$, as it is common in spectroscopy.

\subsection{No--decoherence Hamiltonian}

Considering a small chain of $N$ molecules, the system's Hamiltonian in the site {basis} reads as follows,

\begin{equation}
    \hat{H}_{S}=\sum_{m=0}^{N-1} \epsilon_{m} \Ket{m} \Bra{m} + \sum_{m \neq n}J_{mn}\Ket{m}\Bra{n} 
    \label{rere}
\end{equation}
    
\noindent where $\epsilon_{m}$ is the first excited state energy of the molecule $m$ and $J_{nm}$ is the electronic coupling between the molecules $n$ and $m$. The Hamiltonian (\ref{rere}) for just two molecules (1 qubit), identical to Eq.(\ref{eq:Ham-2}), in the $2 \times 2$ matrix form, reads:
	
	\begin{equation}
	    \hat{H}_{S}=\begin{pmatrix}
	    \epsilon_{0} & J \\
	    J & \epsilon_{1} 
	    \end{pmatrix} \label{diagH} \;.
	\end{equation}

\noindent Its evolution operator is given by

	\begin{equation}
	    \ket {\Psi(t)}=e^{-i\hat{H}_{S}t} \ket {\Psi(0)}\equiv \hat{U}(t)\ket {\Psi(0)} \;.
	\end{equation}

\noindent Although the Hamiltonian \eqref {diagH} possesses non-diagonal elements, finding a good approximation in terms of quantum circuits is relatively straightforward. A possible strategy for this is by finding a diagonalizing transformation, $T$, of the Hamiltonian, such that,

    \begin{equation}
        \hat{H}_{S} = T^{\dag} \hat{H}_{S-diag} T \;.
        \label{xx} 
    \end{equation}

\noindent where $\hat{H}_{S-diag}$ is the diagonal Hamiltonian. Therefore, the evolution operator can be rewritten as follows:

\begin{equation}
    \hat{U}(t)=e^{-i\hat{H}_{S}t} = T^{\dag} e^{-i\hat{H}_{S-diag}t} T\;. \label{xxx}
\end{equation}
    
\noindent The problem now reduces to the approximation of the $T$ operator (and its adjoint) and the Hamiltonian $\hat{H}_{S-diag}$, which can all be efficiently approximated in quantum circuits. The latter operator is diagonal in the site basis, thus the unitary evolution operator can be expressed as

\begin{equation}
    \hat U(t)=e^{-i\hat{H}_{S}t}=T^{\dag}\left[e^{-i\sum_{m=0}^{1}E_{m}t}\right] T=T^{\dag}\left[\prod_{m=0}^{1}e^{-iE_{m}t}\right] T \;.
\end{equation}

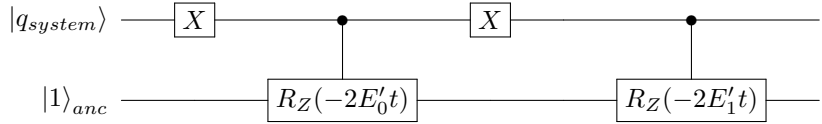
\begin{figure}[H]
       \centering
       \mbox {
       \Qcircuit @C=2em @R=1.6em{ 
        \lstick{\Ket{q_{system}}} & \gate{X} & \ctrl{1} & \gate{X} &\qw & \ctrl{1} & \qw\\
        \lstick{\Ket{1}_{anc}} & \qw & \gate{R_{Z}(-2E^\prime_{0}t)} & \qw &  \qw & \gate{R_{Z}(-2E^\prime_{1}t)} &\qw\\}
       }
       \caption{Implementation of the system's evolution operator. $\Ket{q_{system}}$ is the state vector of the system's qubit in the energy eigenbasis.}
       \label{fig:circuittt}
   \end{figure}

\noindent The $T$ and $T^{\dag}$ matrices can be implemented by simple rotations, $R_{y}(\theta)$ and $R_{y}(-\theta)$, for a two-molecule system. However, for a higher number of molecules, a rotational decomposition algorithm together with the Gray code \cite{nielsen2002quantum}, which decomposes a matrix in the multiplication of a single qubit and CNOT gates, has to be used. Using this particular algorithm the \emph {gate complexity} for $N$ molecules is $\mathcal{O}(N^2 log^{2}[N])$ \cite{nielsen2002quantum}. On the other hand, the diagonalized evolution operator,  

   \begin{equation}
       \hat U(t)=\begin{pmatrix}
       e^{-iE_{0}t}&0\\
       0&e^{-iE_{1}t} \\
       \end{pmatrix}\;,
    \label{oper}
   \end{equation}
   
\noindent translates into trivial phase rotations over each of the energy eigenstates $\Ket{E_{i}}$ of the system with the respective energy eigenvalues $E_{i}$. This operator can be constructed as a sequence of $CR_{Z}(\phi_{i})$ gates applied to an \emph {ancilla qubit} (initialized at $\Ket{1}$), where the angle is given by $\phi_{i}= - 2E_{i}t$, $i=1,2$. The $X$ gates are used to "select" the eigenvector to which the controlled rotation is to be applied. The circuit implementation of the operator defined in \eqref{oper} is illustrated in Figure \ref{fig:circuittt}. The gate complexity of this operator, in terms of single qubit and CNOT gates for $N$ molecules, is $\mathcal{O}(N\log[N])$. 
   
\noindent For the whole circuit, resulting from the sequencing of $T^{\dagger} \hat{H}_{S-diag} T$, the number of qubits required to simulate a molecular chain of $N$ elements is $2\log _2 N$ and the gate count scales with $\mathcal{O}(N^2 \log^{2}_2 N)$ single qubit and CNOT gates. The transformations $T$ and $T^{\dag}$, in the general case, possess a high circuit depth, which makes the system hard to simulate accurately, with low error rate, in the current available quantum computers. 

\subsection{Introducing decoherence into the system}
\label{sec:decoherence}

\begin{figure}[H]
        \centering
        \includegraphics[width=0.6\linewidth]{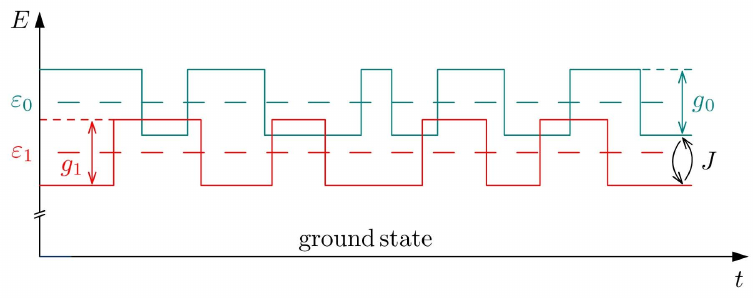}
        \caption{Uncorrelated random fluctuations applied to donor and aceptor's excited state energies, $\epsilon_{0}$ and $\epsilon_{1}$. Each molecule is affected by one fluctuator, which generates a telegraph-type classical noise. The fluctuators switch randomly between the positive and negative value at a given fixed rate, so that the periods of time when the molecule energy is constant,  $\epsilon_{m}+g_m/2$ or $\epsilon_{m}-g_m/2$, are random.  
        $J$ is the coupling strength between the molecules that can be seen as the rate of hoppings between these fluctuating energy levels.}
        \label{fig:energy_}
    \end{figure}

We shall implement artificial decoherence as pure-dephasing by adding Markovian fluctuations to the Hamiltonian. This approach is considered a good approximation in the \emph{high-temperature} regime for the bath \cite{leegwater1996coherent,rebentrost2009environment,breuer2002theory}. The actual algorithm to be used is the one of \cite{wang2011quantum}, which is used to simulate open quantum systems, with pure dephasing, modeling the action of the decoherence as classical random fluctuations (a telegraph-type classical noise affecting the system). The actual Hamiltonian for this system reads as

\begin{equation}
    \hat{H} = \hat{H}_{S}+\hat{H}_{F}
\end{equation}

\noindent and it consists of the system Hamiltonian, $\hat{H}_{S}$, of the previous section and the perturbation of a \emph {bi-stable fluctuator} environment, $\hat{H}_{F}$. The latter simply shifts the energy by a constant value for each molecule, $\pm g_m/2$, as illustrated in Fig. \ref{fig:energy_}. Explicitly,
\begin{equation}
    \hat{H}_{F}=\sum_{m=0}^{1}\chi_{m}(t)\hat{A}_{m}
\end{equation}
where $\hat{A}_{m}\Ket{m}\Bra{m}$ is the {projection operator} and considering one fluctuator interacting with each molecule $m$,
\begin{equation}
    \chi_{m}(t) = g_{m} \xi_{m}(t) \;.
\end{equation}
 {The function $\xi_{m}(t)$ switches the fluctuator between the positive and negative values (appearing randomly) at a given fixed rate $\gamma$} and $g_{m}$ is the fluctuation strength (or the coupling strength to a molecule $m$).
Physically, the action of the fluctuations is typically stronger for the excited states \cite{leegwater1996coherent,adolphs2006proteins} and $g$ can be larger than the donor-acceptor coupling $J$.

\noindent The implementation of such random bi-valued function $\xi_{m}(t)$, can be done in a straightforward way by a classical pseudo-random numbers generator with a probability of $50\%$ of the values $-1/2$ and $1/2$. For circuit generation purposes, the values resulting from the random sampling have to be provided in advance of the quantum simulation. 
    
The fluctuator interaction Hamiltonian and the system Hamiltonian do not commute, so, in order to generate an appropriate quantum circuit, one needs to apply an approximation technique such as the Trotter product formula \cite{wiebe2010higher}. Under this approximation, the unitary evolution operator of the Hamiltonian, for a time $t=N_i\Delta t$, where $N_i$ is the number of iterations and $\Delta t$ is the iteration time-step, becomes

\begin{equation}
    U(N_i \Delta t)=\left( e^{-i\hat{H} \Delta t}\right)^{N_i} = \left( e^{-i\hat{H}_{F}\Delta t}T^{\dag}e^{-i\hat{H}_{S}\Delta t}T \right)^{N_i}=\left( \left[\prod_{m=0}^{1}e^{\pm i\frac{g_{m}}{2}\Delta t}\right]T^{\dag}\left[\prod_{m=0}^{1}e^{-iE_{m}\Delta t}\right]T\right)^{N_i} \,,
    \label{eq:UN}
\end{equation}
 where $E_{m}$ denote the eigenvalues of the system Hamiltonian. 
 
\begin{figure}[H]
    \centering
       \mbox{
       \Qcircuit @C=0.5em @R=1.5em{ 
        \lstick{\Ket{q_{system}}} & \gate{R_{y}(\theta)} & \gate{X} & \ctrl{1} & \gate{X} &\qw & \ctrl{1} & \qw & \gate{R_{y}(-\theta)} & \qw & \gate{X} & \ctrl{1} & \gate{X} &\qw & \ctrl{1} & \qw &\\
        \lstick{\Ket{1}_{anc}} & \qw & \qw & \gate{R_{Z}(-2E'_{0}\Delta t)} & \qw &  \qw & \gate{R_{Z}(-2E'_{1}\Delta t)} &\qw & \qw & \qw & \qw & \gate{R_{Z}(\pm g' \Delta t)} & \qw &  \qw & \gate{R_{Z}(\pm g'\Delta t)} & \qw \\}
       }
       \caption{Implementation of one iteration of the system with decoherence algorithm. Here $\Ket{q_{system}}$ represents the system's qubit state vector in the site basis.}
       \label{fig:circuitt}
   \end{figure}
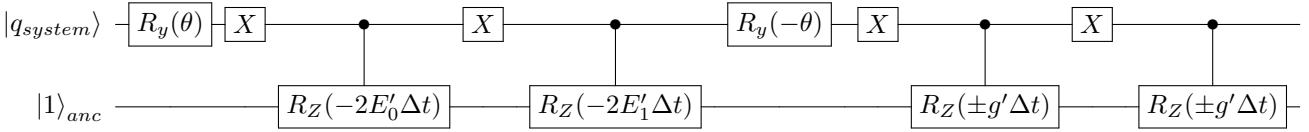

 Note that the projection operator $\hat{A}_{m}$ is not present in the evolution operator (\ref{eq:UN}) because the latter is used in its eigenbasis, i.e. the site basis. The fluctuator interaction evolution operator $e^{\pm i \frac{g_{m}}{2}\Delta t}$ is a \emph {selective rotational gate} over a molecule $m$ ($\ket {m}$), which can be implemented by a set of X gates and a controlled gate $CR_{Z} (\phi_{m})$ with angle $\phi_{m}=\pm g_{m}\Delta t$,
applied over an \emph {ancilla} qubit initialized at $\Ket{1}$. The whole circuit is presented in Fig. \ref{fig:circuitt} for one iteration. The fluctuator waiting time (interval of time between switches), i.e. $\frac{1}{\gamma}$, can only be equal or higher than the iteration time-step, $\Delta t$. The switching in the fluctuator-molecule coupling strength is performed at every $\frac{1}{\gamma \Delta t}$ iterations, where $a\Delta t =  \frac{1}{\gamma}\text{ },\; a \in \mathbb{N} $.

Usually in the study of open quantum systems with a dilated system's Hilbert space (as is the case here), different measurement techniques are required \cite{nielsen2002quantum,wang2011quantum}, however in this case, the open system is simulated in a closed form so,  similarly to the \emph {no decoherence} case, the measurement over the site basis suffices. The full algorithm (random values generator plus the actual simulation) must be performed several times, so that the results of all runs are averaged.
    
Let us consider the simulation for a time $t$, using an iteration time step $\Delta t$, and assuming that the environment can have more than one fluctuator interacting with each molecule as well as the chain can have more than just two elements. Then the fluctuator interaction evolution operator requires the following gate resource complexity for a single run: 
\noindent $\mathcal{O}(\frac{t}{\Delta t}[N(\log_{2}N+F)])$ single qubit and CNOT gates, where $N$ is the number of molecules and $F$ is the number of fluctuators interacting with each one. 
    
In the implementation of the system with decoherence, the algorithm gate resources complexity is $\mathcal{O}(\frac{t}{\Delta t}[N^2 $ $ \times \log_2^{2}N+NF])$ for a single run. This simulation, yet again, possess a very high circuit depth which makes its application unfeasible in quantum computers.
The number of necessary qubits is the same as in the no decoherence simulation ($2\log_2 N$). 

It also requires $\mathcal{O}(N R\sum_{j=0}^{F} t\gamma_{j})$ random numbers to be classically generated, where $R$ is the number of runs of the algorithm and $\gamma_{j}$ is the switching rate of the fluctuator $j$ interacting with the molecule.
The number of required simulation runs to average the results and obtain an error $\epsilon>0$, is predicted to scale as $\mathcal{O}\left({[F\frac{t}{\Delta t}]^{2}}/{\epsilon^{2}}\right)$. This complexity is calculated based on the possible non-degenerate energy state outcomes of the entire chain in the simulation for a time $t$. These outcomes are caused by the bi-stable random fluctuations, therefore, the possible non-degenerate energy state outcomes for each molecule obey a discrete Gaussian probability distribution. 

\section*{Results}\label {results}

We conducted simulation experiments for the quantum transport in a molecular chain using the algorithm described in the previous section. We executed the simulation for the coherent system on a \emph {real} quantum computer, the IBM Q of 5 qubits, while the \emph {pure dephasing} scenario was simulated on the \emph{QASM} quantum simulator, both in the near-resonant and non-resonant regimes. For the validation purposes, we compared the results for the coherent system with the theoretical predictions obtained by solving the Schr\"{o}dinger equation (see {Supplementary Information A.2}). 

As for the \emph {decoherent} regime, we used a classical computation of the stochastic Haken-Str\"{o}bl model \cite{haken1973exactly,rebentrost2009environment}. The simulations and circuits involved, encoded in the Qiskit platform \cite{cross2018ibm}, can be performed in the following url:  \url{https://github.com/jakumin/Photosynthesis-quantum-simulation}.

\subsection{Coherent regime}

The scenario for this regime was simulated with a simple chain of two molecules. As discussed in section \textit{Materials and Methods} and using the parameters as proposed in \cite{wang2018efficient}, we define the system's Hamiltonian as follows: 

(Near-resonant regime)
\begin{equation}
	        H_{S}=\begin{pmatrix}
	        13000 & 126 \\
	        126 & 12900 \\
	        \end{pmatrix} \label{hsa} cm^{-1}\; ;
	\end{equation}
	
\begin{figure}[H]
        \centering
        \includegraphics[width=0.7\linewidth]{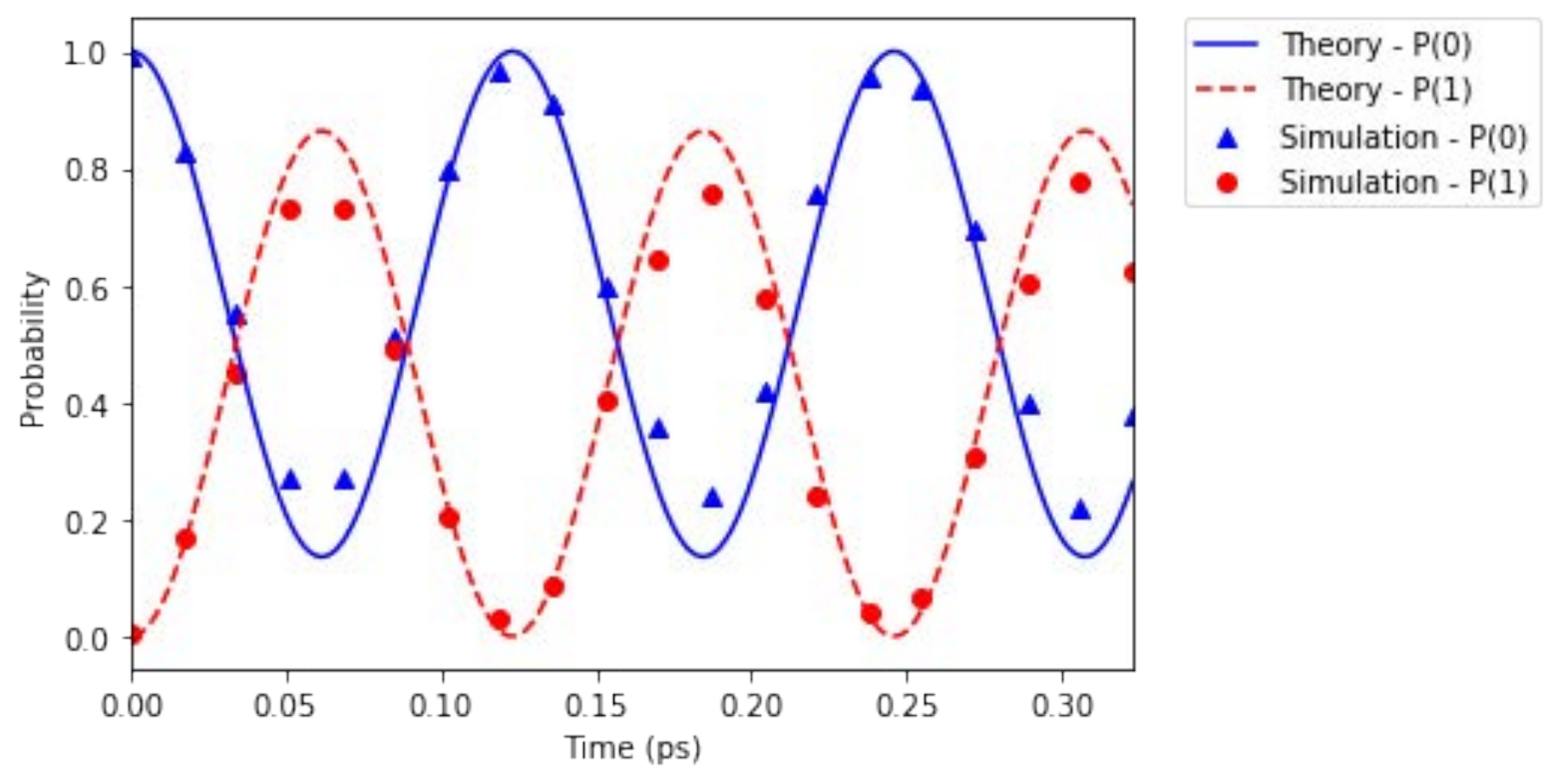}
        \caption{Evolution dynamics of the isolated system obtained by employing the quantum algorithm for the near-resonant system: simulation results (points) and theory (lines).}
        \label{fig:res}
    \end{figure}

\begin{figure}[H]
        \centering
        \includegraphics[width=0.7\linewidth]{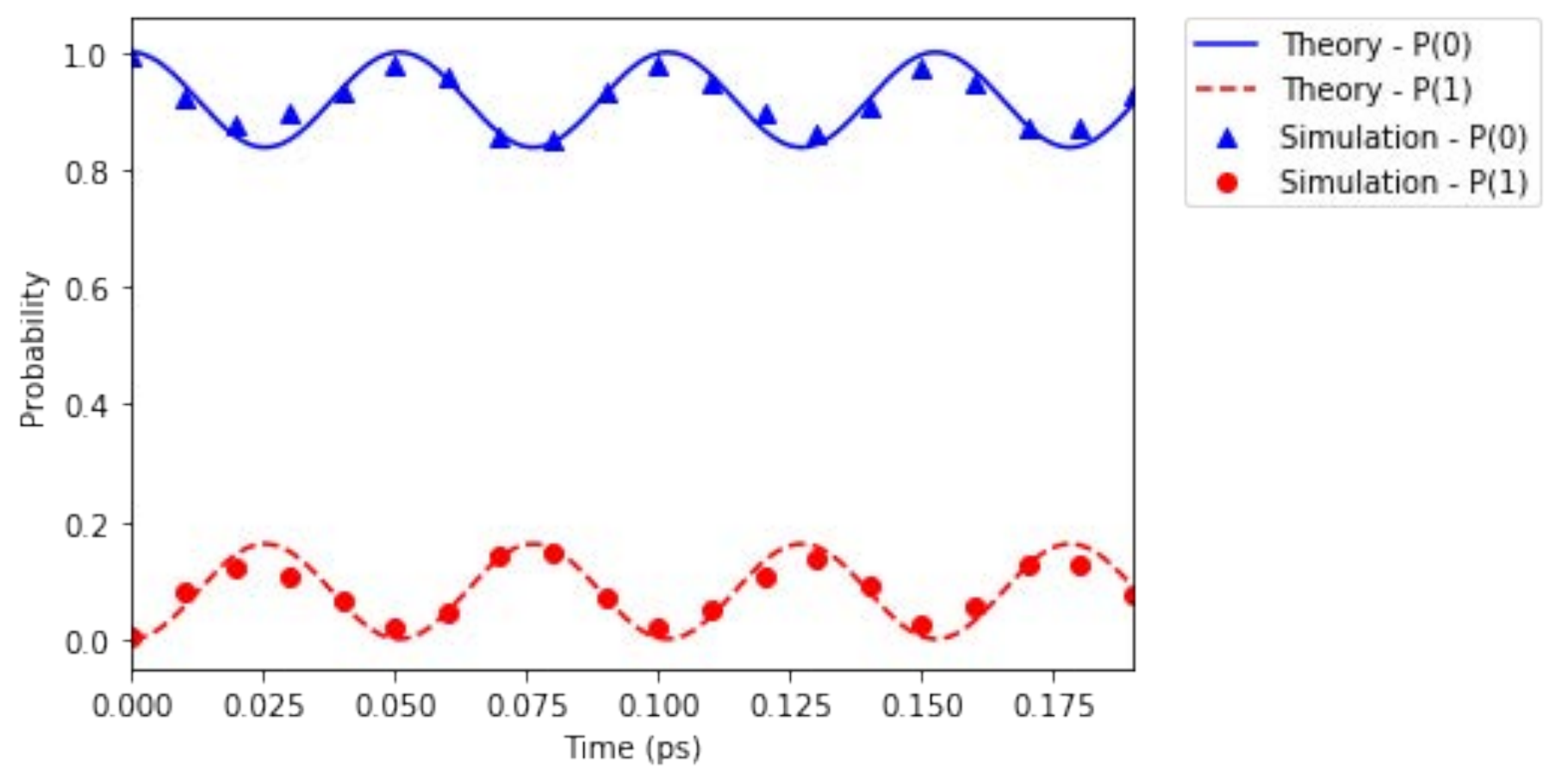}
        \caption{Evolution dynamics of the isolated system obtained by employing the quantum algorithm for the non-resonant system: simulation results (points) and theory (lines).}
        \label{fig:nres}
    \end{figure}

(Non-resonant regime)
\begin{equation}
	        H_{S}=\begin{pmatrix}
	        12900 & 132 \\
	        132 & 12300 \\
	        \end{pmatrix} \label{hsb} cm^{-1}\; .
	\end{equation}
The results for both regimes were obtained using an actual quantum device (the IBMQ london of 5 qubits) and can be seen in Figs. \ref{fig:res} and \ref{fig:nres}, respectively. Due to the stochastic nature of quantum computers, the experiments were conducted with $2048$ shots for each time value. The specific optimized quantum circuits used in this experiment are presented in {Supplementary information A.3}. In the following results, the probability of the donor and acceptor molecules being excited is denoted by $P(0)=\bra{0}\rho_{S}(t)\ket{0}$ and $P(1)=\bra{1}\rho_{S}(t)\ket{1}$, respectively. 
Taking the fluctuator's switching rate to be $\gamma = 0$ or the fluctuator-molecule coupling strength to be $g=0$, one has the coherent regime. These simulations show the limiting case of the Redfield regime, i.e. the very weak  system-environment coupling, $g\ll J$. The quantum beatings, observed in the simulation results, can be thought of as a \emph{reversible} transfer of energy between the molecules, where the excitation goes back and forth across the molecules \cite{cheng2007elucidation}.

In the performed simulations, the near-resonant and non-resonant regimes have a maximum probability of $\sim 90\%$ and $\sim 20\%$, respectively, of the energy being transferred to the acceptor molecule. Using the quantum Liouville equation \cite{mohseni2014quantum} (see {Supplementary information A.2}), the period of the quantum beating is $T_{near-res}\approx123$ $fs$ for the near-resonant regime and $T_{non-res}\approx51$ $fs$ for the non-resonant regime. These periods are in the femtosecond timescale of the experimentally observable quantum beatings \cite{engel2007evidence,collini2010coherently,panitchayangkoon2010long}. The simulation results show a similar behaviour as those predicted by the {Schr\"{o}dinger} and {quantum Liouville} equations, where the off curve points are predominantly originated by errors in the quantum hardware.

\subsection{Decoherent regime}

The scenario for the regime with decoherence introduced is, in some respect, similar to the one presented for the coherent regime for a chain of two molecules. No further changes are made to the Hamiltonian discussed in the section \textit{Introduction of decoherence in the system}. The quantum simulation results are compared with a theoretical evolution based on the stochastic Haken-Str\"{o}bl model, in the form of the Lindbland master equation \cite{haken1973exactly,rebentrost2009environment}. The Lindbland equations were solved in a classical computer using \emph{Qutip} \cite{johansson2012qutip}, a \emph {quantum open systems software framework}. The set of Lindbland equations, correspondent to the model in this setting, had one free parameter regarding the {environment}, the \emph {dephasing rate}, $\gamma_{deph}$. The Lindbland equation in the Haken-Str\"{o}bl model reads: 
\begin{equation}
    \frac{d\rho}{dt} = \mathcal{L}[\rho] = -i [H_{S},\rho] + \gamma_{deph} \sum_{m} (L_{m}\rho(t)L_{m}^{\dag} - \frac{1}{2} \rho(t)L_{m}^{\dag} L_{m} - \frac{1}{2} L_{m}^{\dag} L_{m}\rho(t) )\label{lind}
\end{equation}
where $L_{m} = \Ket{m}\Bra{m}$ are the Lindbland operators, responsible for the system-environment interaction. The system Hamiltonian, $H_{S}$, is given by the matrix \eqref{hsa} for the near-resonant system and the matrix \eqref{hsb} for the non-resonant system.

For each quantum simulation performed, a fitting process has been employed by adjusting the dephasing rate of the Haken-Str\"{o}bl model, so that the system's evolution in both classical and quantum algorithms have similar behaviours. This enables one to perform a direct comparison between both theories and to find the actual dephasing rate of the modeled environment over the various regimes considered in this work.

\begin{figure}[H]
        \begin{subfigure}{.5\textwidth}
        \centering
   \caption{$g=100$ $cm^{-1}$, $\gamma_{deph}=2.3$ $THz$.}
  \includegraphics[width=\linewidth]{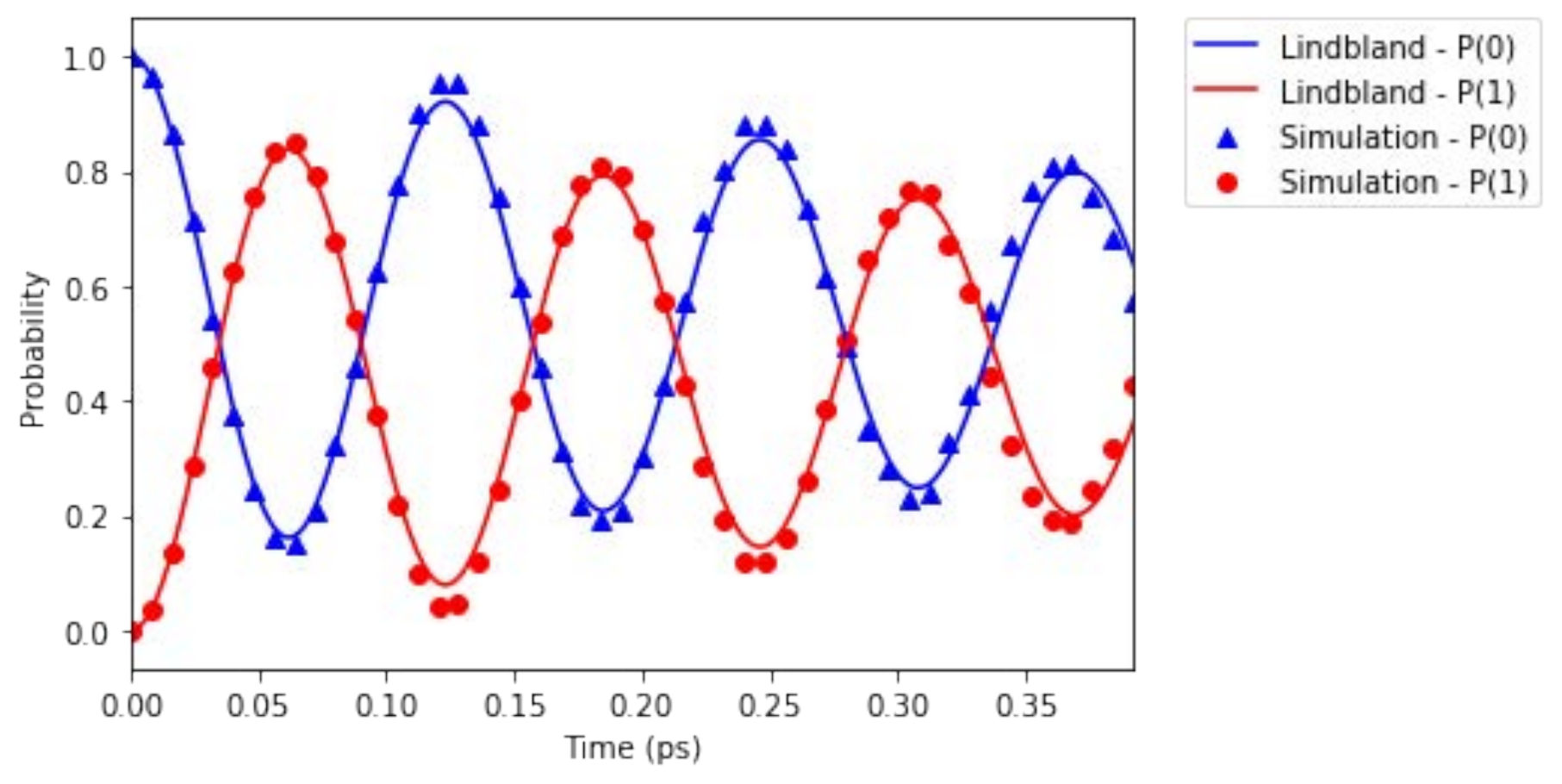}
  \label{resred1}
\end{subfigure}
\begin{subfigure}{.5\textwidth}
\centering
   \caption{$g=300$ $cm^{-1}$, $\gamma_{deph}=10$ $THz$.}
  \includegraphics[width=\linewidth]{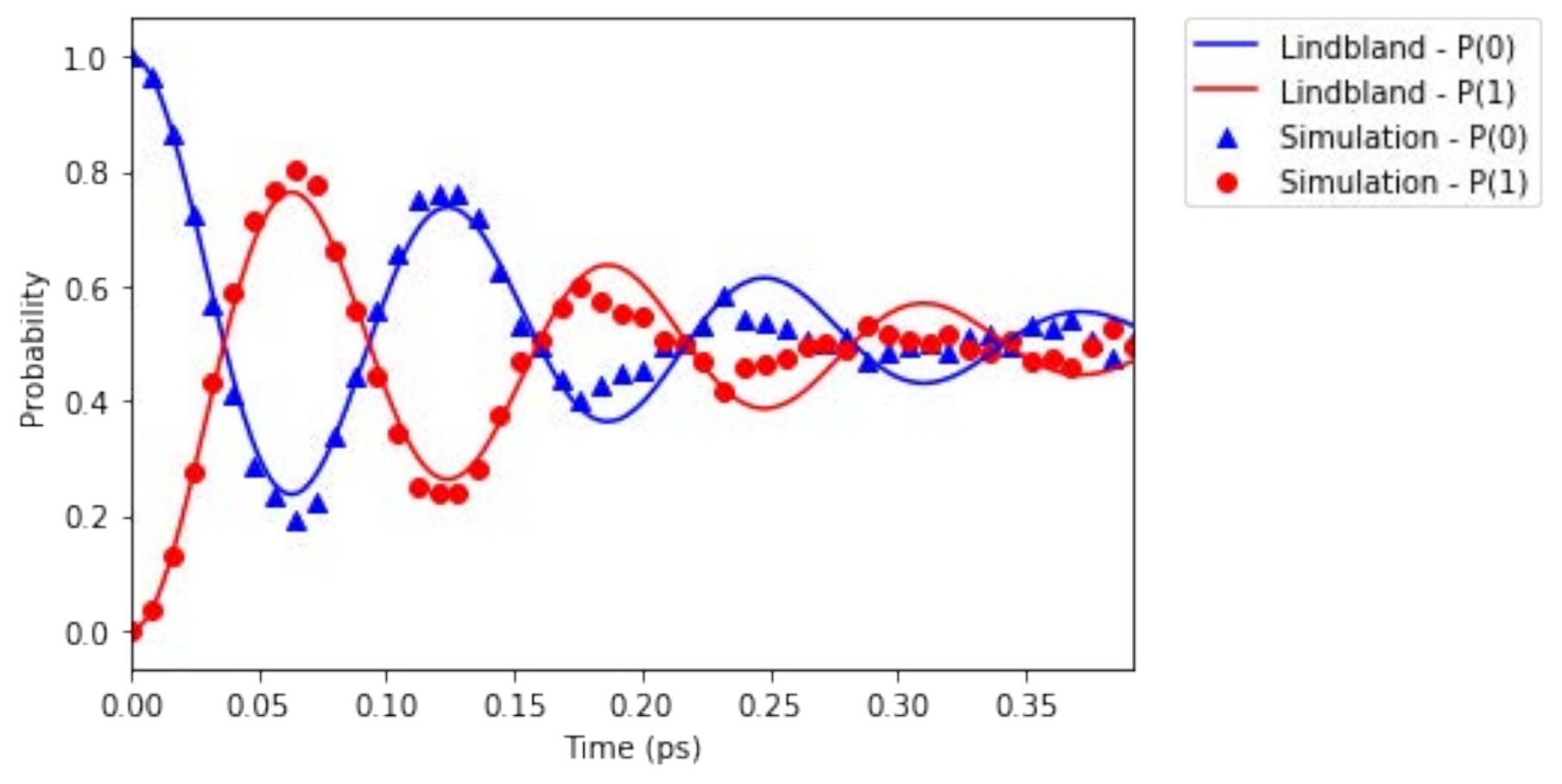}
  \label{resred2}
\end{subfigure}
\begin{subfigure}{.5\textwidth}
\centering
   \caption{$g=700$ $cm^{-1}$, $\gamma_{deph}=41$ $THz$.}
  \includegraphics[width=\linewidth]{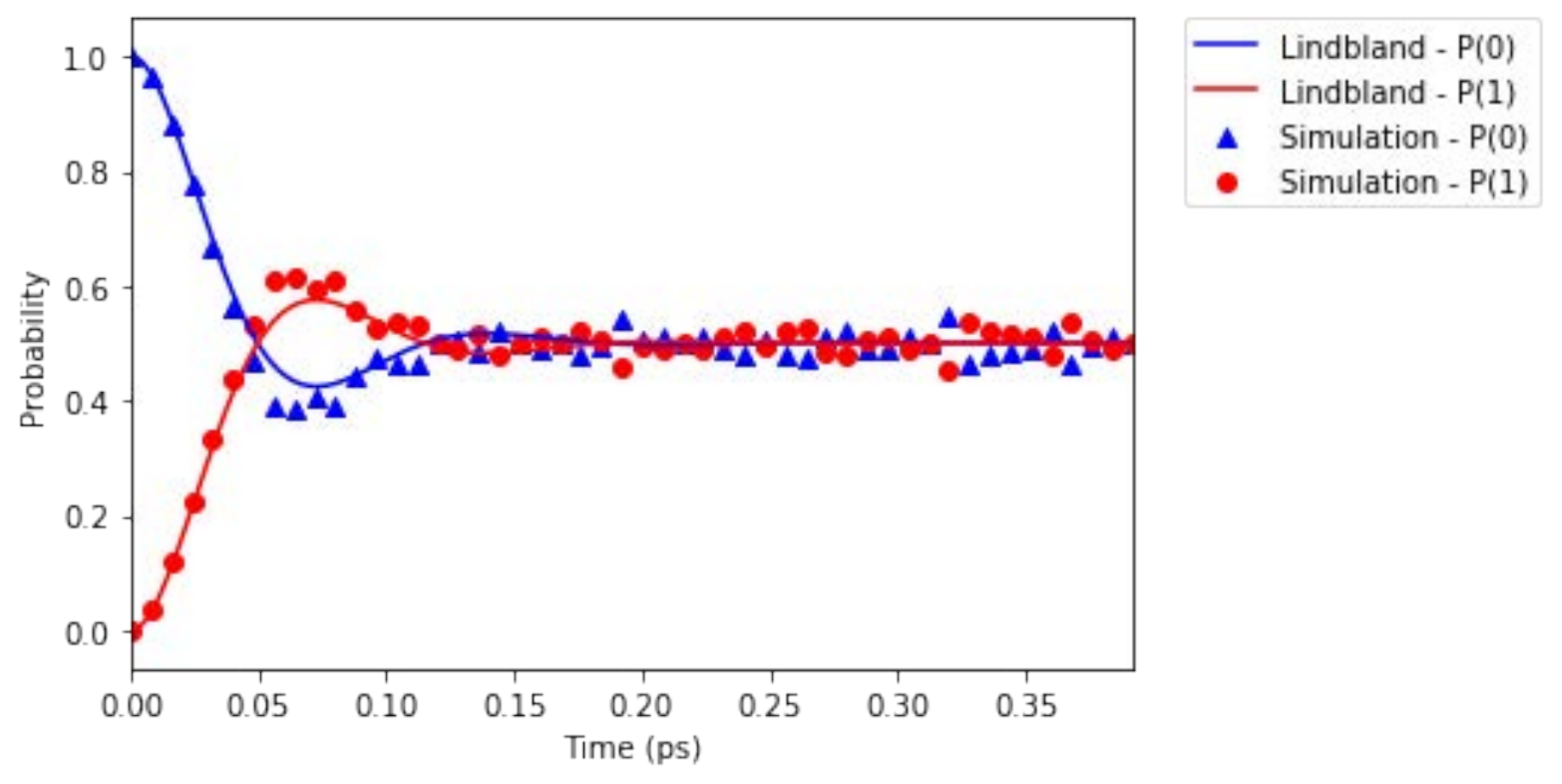}
  \label{resred3}
\end{subfigure}
\begin{subfigure}{.5\textwidth}
\centering
   \caption{$g=1000$ $cm^{-1}$, $\gamma_{deph}=70$ $THz$.}
  \includegraphics[width=\linewidth]{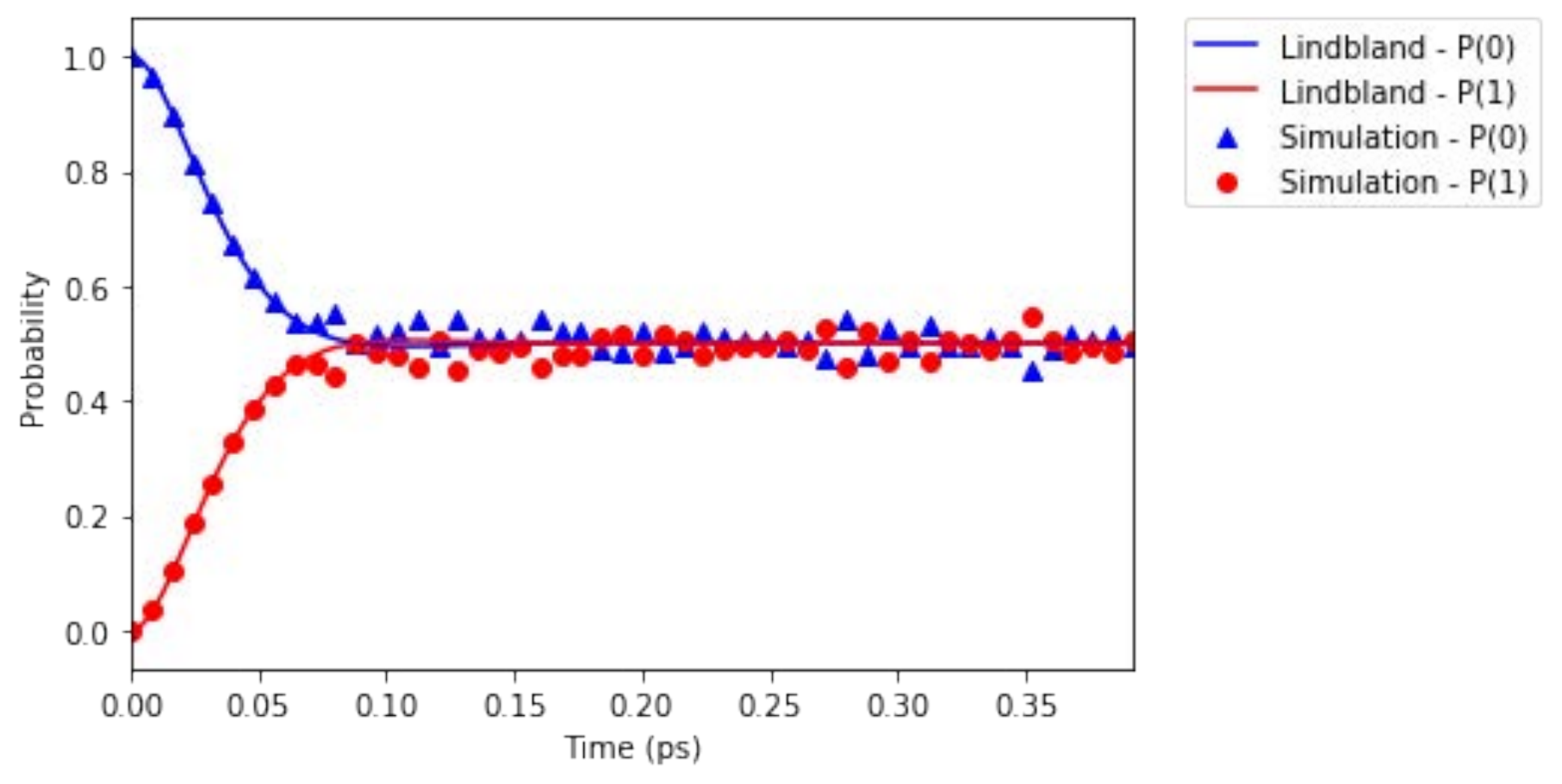}
  \label{resred4}
\end{subfigure}
\caption{Evolution dynamics of the system with decoherence obtained by employing the quantum algorithm for the near-resonant system: simulation results (points) and theory (lines).}
\label{reslind}
\end{figure}

\begin{figure}[H]
        \begin{subfigure}{.5\textwidth}
        \centering
   \caption{$g=100$ $cm^{-1}$, $\gamma_{deph}=2.3$ $THz$.}
  \includegraphics[width=\linewidth]{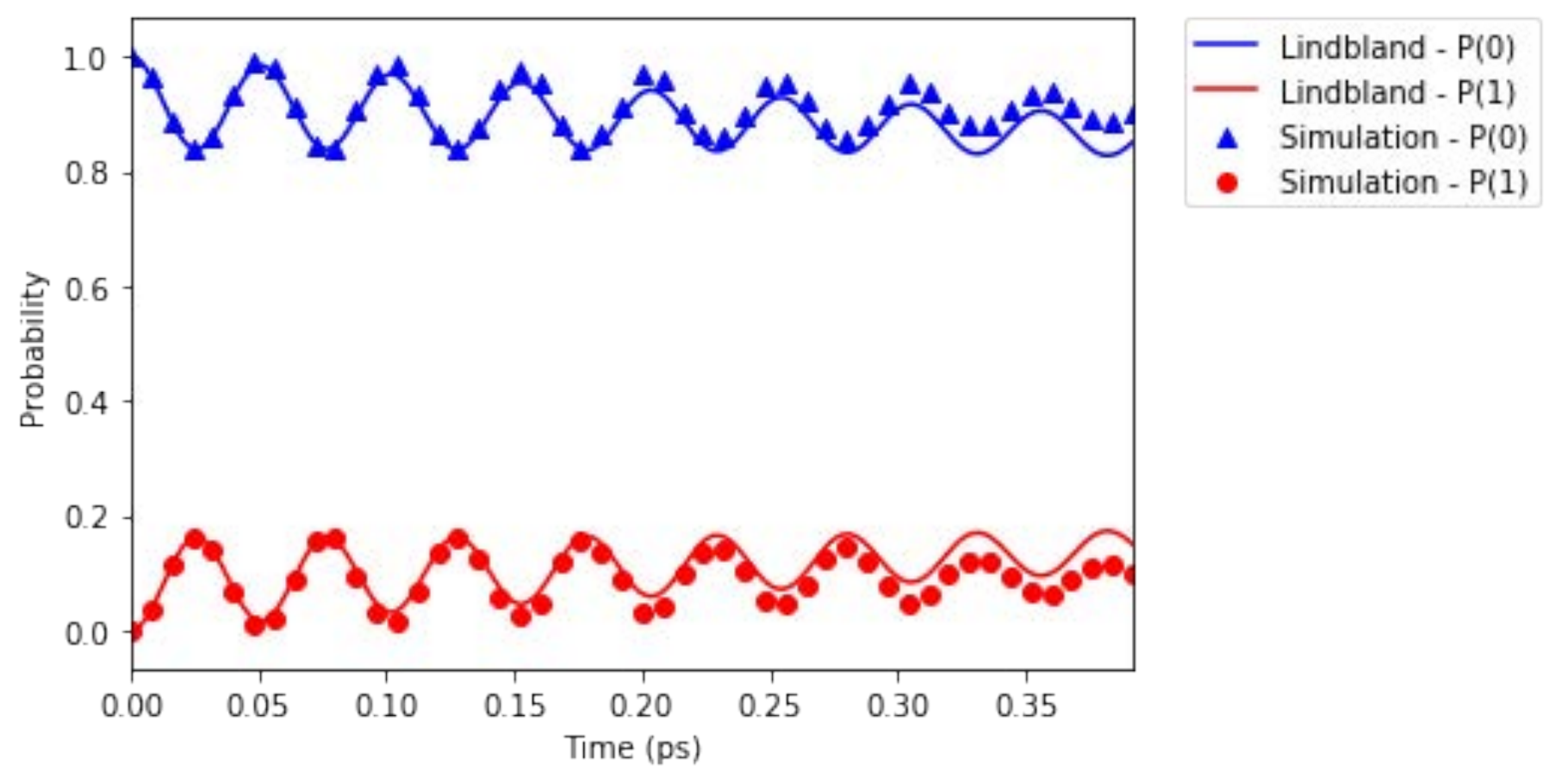}
  \label{nresred1}
\end{subfigure}
\begin{subfigure}{.5\textwidth}
\centering
   \caption{$g=300$ $cm^{-1}$, $\gamma_{deph}=10$ $THz$.}
  \includegraphics[width=\linewidth]{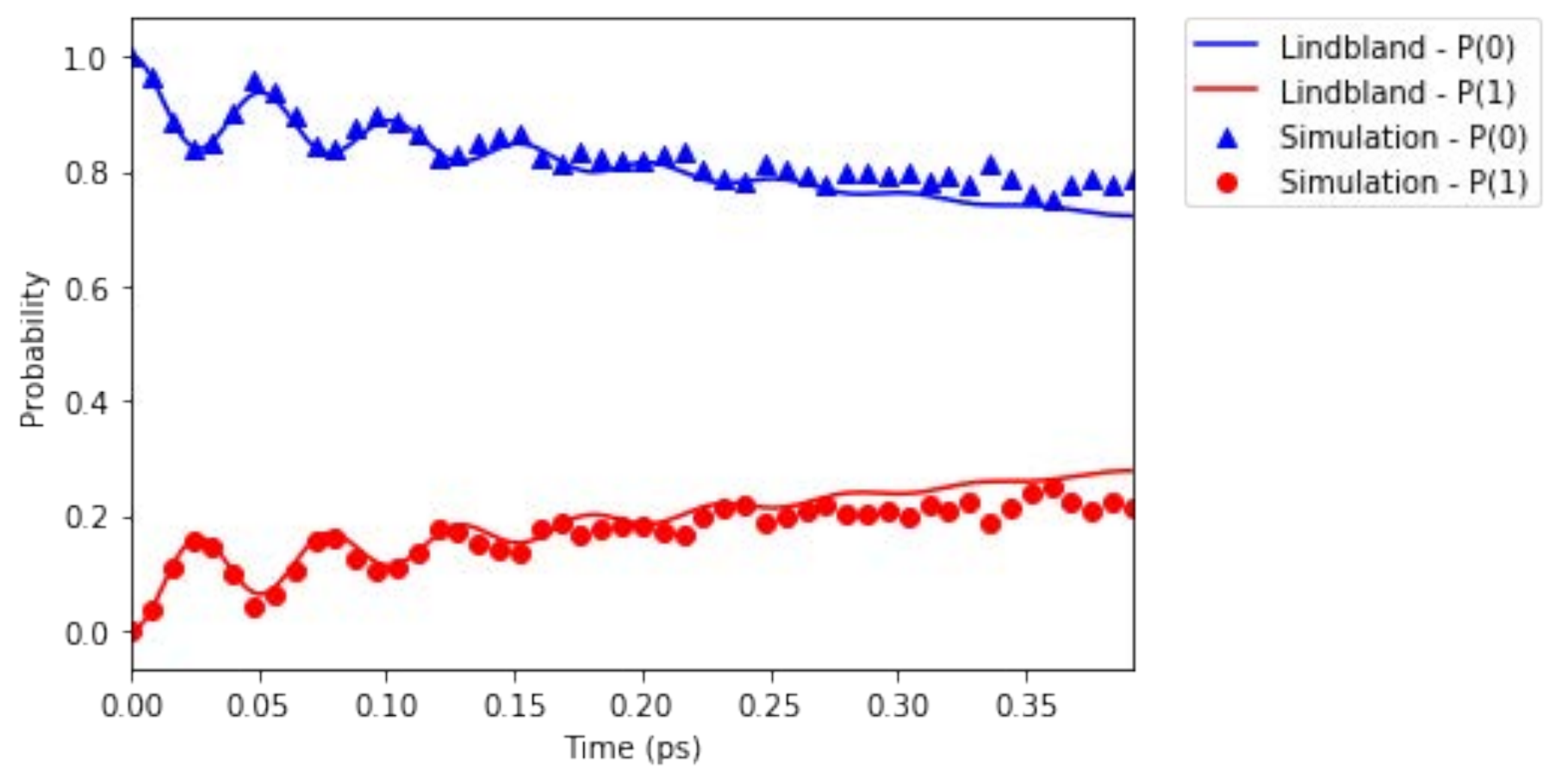}
  \label{nresred2}
\end{subfigure}
\begin{subfigure}{.5\textwidth}
\centering
   \caption{$g=700$ $cm^{-1}$, $\gamma_{deph}=41$ $THz$.}
  \includegraphics[width=\linewidth]{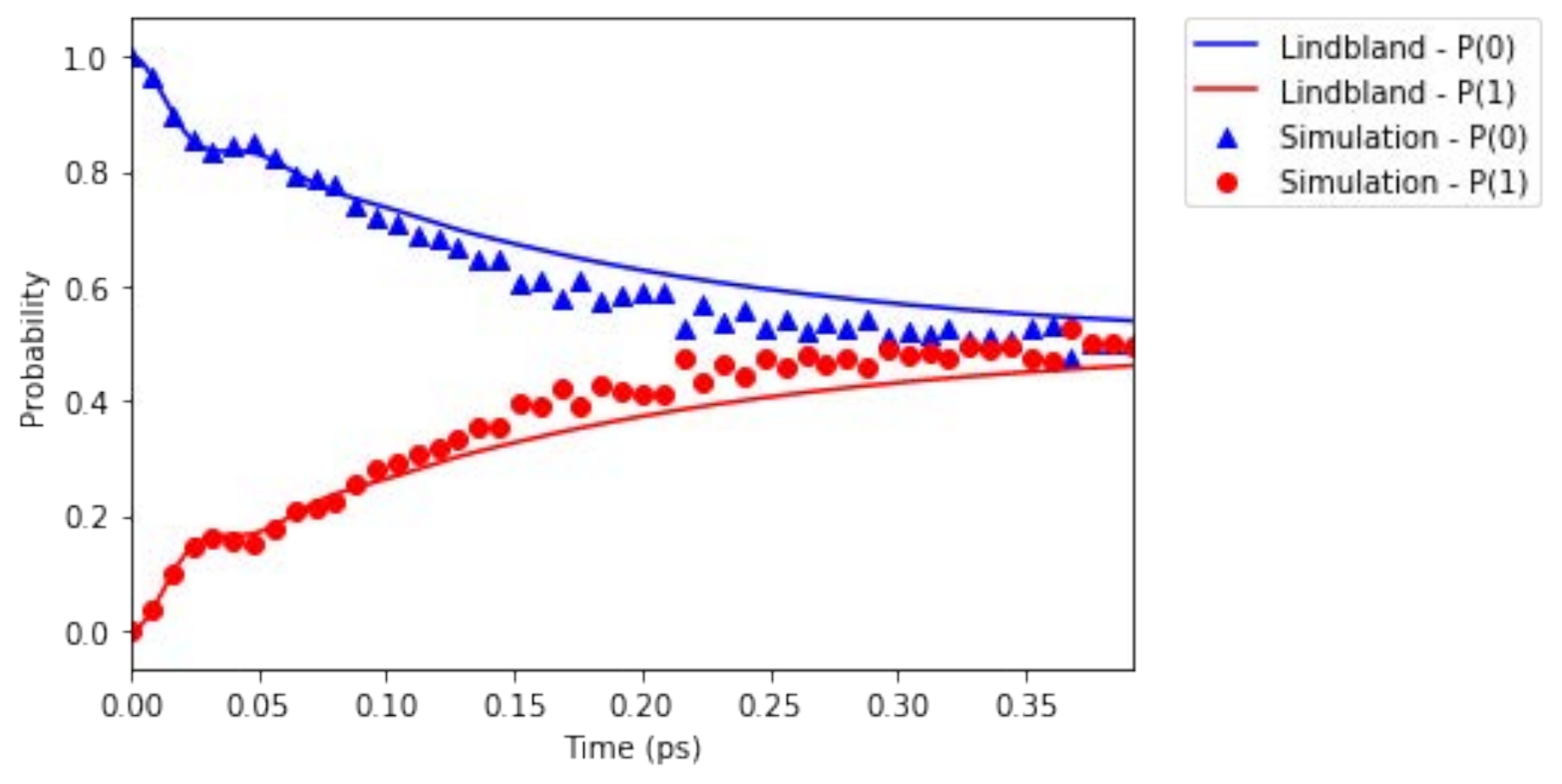}
  \label{nresred3}
\end{subfigure}
\begin{subfigure}{.5\textwidth}
\centering
   \caption{$g=1000$ $cm^{-1}$, $\gamma_{deph}=70$ $THz$.}
  \includegraphics[width=\linewidth]{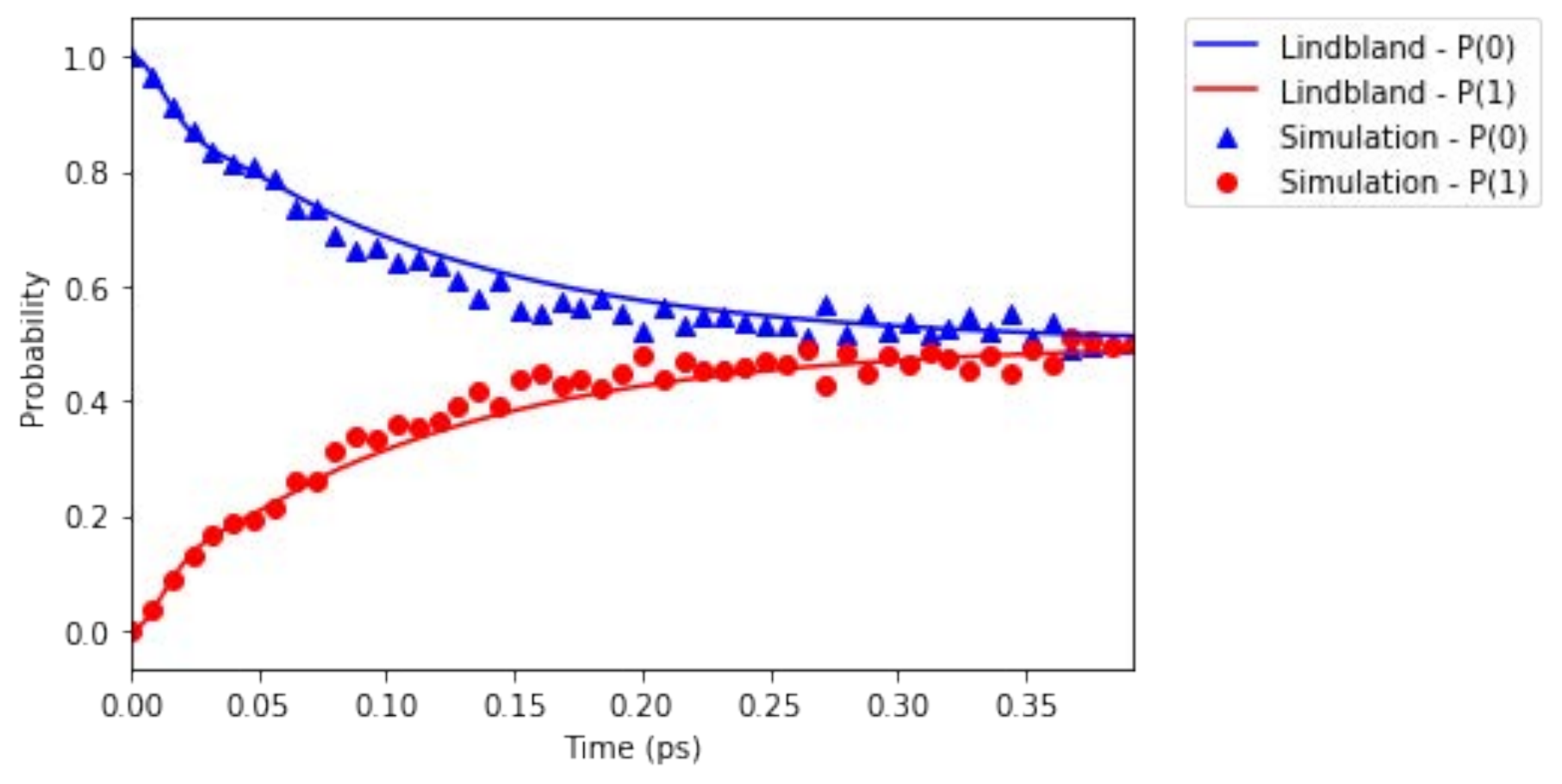}
  \label{nresred4}
\end{subfigure}
\caption{Evolution dynamics of the system with decoherence obtained by employing the quantum algorithm for the non-resonant system: simulation results (points) and theory (lines).}
\label{nreslind}
\end{figure}

The environment contains only one fluctuator interacting with each molecule with switching rate $\gamma=125$ $\text{THz}$. As mentioned above, the dephasing rate, $\gamma_{deph}$, for the Lindbland equation is adjusted to the behaviour of the system under the action of a fluctuation strength $g$. For a range of fluctuation strengths of $[100,1000]$ $cm^{-1}$, in the quantum algorithm, and the corresponding dephasing rate of the Haken-Str\"{o}bl model lies in the $\sim [2.3,70]$ $\text {THz}$ range. Due to the existence of random fluctuations, large number of samples had to be generated. The algorithm was implemented with $250$ runs, where $5000$ shots were performed for each time $t$. Figures \ref{reslind} and \ref{nreslind} present the simulation results for different values of the fluctuation strength, along with the theoretical evolution dynamics, for the near-resonant and non-resonant systems, respectively.

\noindent It is seen in Figures \ref{reslind} and \ref{nreslind} that oscillation amplitudes decay over time, as expected, due to the loss of relative phase coherence between the excited states of the two molecules, evidenced by the disappearance of the quantum beatings. This is associated with the irreversible evolution when the system loses its capacity of performing \emph {coherent} transport. Additionally, it is clear that the system is led to a classical distribution of the populations in the {site eigenbasis}.  

In the regime under the study, where the environment is assumed to be at thermal equilibrium, the final probability distribution is calculated in the limit of the classical {Boltzmann} distribution $\Bra{m}\rho_{S}(t \to \infty)\Ket{m}= const \times e^{-\frac{ \epsilon_{m}}{k_{B}T}}$. Here $k_{B}$ is the Boltzmann constant, $T$ is the temperature of the bath and $const$ is a normalization constant \cite{breuer2002theory}. Taking the limit of very high temperatures, the population terms approach the Boltzmann distribution $\Bra{0}\rho_{S}(t \to \infty)\Ket{0}\approx \Bra{1}\rho_{S}(t \to \infty)\Ket{1}\approx \frac{1}{2}$, which is compatible with the results obtained.  The relaxation can not be fully observed in Figs. \ref{resred1}, \ref{nresred1} and \ref{nresred2} because a very large number of iterations would be required for this.

 The switching rate must be high enough to observe the dephasing effects. Here we used a value $\approx 33$ times larger than the transfer rate,  $J$ (that is, the fluctuator waiting time must be shorter than $J^{-1}$). As observed in the simulations, it is a suitable value for observing the relevant effects of random fluctuations in the system. At very low rates, it leads the system's evolution to a behaviour similar to the previously observed in the no-decoherence regime, Figs. \ref{fig:res} and \ref{fig:nres}.

The time that coherence lasts in the system is essentially defined by the fluctuation strength, $g$: in Figs. \ref{resred1}, \ref{resred2}, \ref{nresred1} and \ref{nresred2} (lower $g$) the coherence is maintained for some time, while in Figures \ref{resred3}, \ref{resred4}, \ref{nresred3} and \ref{nresred4} (higher $g$) it is quickly suppressed. In the latter regime, an approximated diffusive motion drives the system's evolution, where quantum beating is practically absent.
The time that the quantum beating lasts in these simulations (until it reaches an approximate non-oscillating behaviour), is about $350$ $fs$ in Figure \ref{resred2} (near-resonant system) and $200$ $fs$ in Figure \ref{nresred2} (non-resonant system), with a fluctuation strength $g=300$ $cm^{-1}$. At a longer time, it has been experimentally observed to persist ($t > 660$ $fs$ \cite{panitchayangkoon2010long}), a timescale which could be modeled in the present simulation by changing the environment parameters, i.e. lowering the fluctuation strength $g$ as can be observed in Figures \ref{resred1} and \ref{nresred1}.

\section*{Discussion and Conclusions} 

Two main conclusions can be drawn from the presented results: 
\begin {itemize}
\item There is a very good agreement between the solution of the Schr\"{o}dinger equation and the \emph {coherent} quantum algorithm results in the reproduction of the purely oscillatory evolution of the isolated quantum system.
\item There also is a good agreement between the results obtained by the Haken-Str\"{o}bl model and the quantum algorithm. The increase of the {dephasing rate} imply an increase in the fluctuation strengths thus, a \emph {faster} suppression of the {quantum beatings} can be observed, as  predicted theoretically \cite{rebentrost2009environment}.   
\end {itemize}
\noindent Therefore, the correctness of the results obtained in the quantum simulations is verified.

The results obtained in the present work were not directly compared with Ref. \cite{wang2018efficient}, due to the different timescales used. The major difference lies in the physical implementation, NMR\emph { vs} universal quantum computer, where there might be an advantage for the former from the point view of the scalability and reliability, at the current state of quantum technology. However, there is a clear advantage of the quantum computer, from the point of easiness of implementation, as it is also possible to implement circuits of arbitrary precision, harder to do with the NMR simulator, which is dependent on a {Hamiltonian mapping} process. The computational advantage verified for the {NMR} simulator still holds after the present work, as the number of executions for the algorithm of this work is polynomial on the precision required, although the circuit generation maybe problematic, as a matrix diagonalization operation is necessary (complexity estimated in $O(N
^{3})$).

To conclude, we proposed a quantum algorithm to simulate the energy transfer phenomenon present in general photosynthesis, under the presence of {quantum coherence} between the molecules and the {decoherence} effects caused by environmental interference. Using this algorithm we also performed simulations in the {commercially} available quantum computer of IBM, the IBM Q of 5 qubits for the {coherent scenario} and in the {quantum simulator} (QASM) for the {decoherent scenario}. 
For validation purposes, we also computed the evolution of analogous systems using well-established (classical) methods in literature, obtaining quite similar results between the methods. The results obtained were also in agreement with the predictions that can be found in literature, for the role of the quantum coherent and dephasing effects in the energy transport of photosynthesis: for the high temperature environment here defined, it was clear that dephasing, modelled as energy fluctuations in the site energies, limited the time quantum coherence lasts in light-harvesting antenna. Moreover, it was also verified that the fluctuation strength and the switching rate of the Markovian {fluctuator} environment are directly related with the energy transfer efficiency, allowing the simulation of different transport regimes by setting them appropriately. 

Similar to Ref. \cite{wang2018efficient}, this setting revealed itself as an interesting platform for the study the quantum and environmental effects in a small photosynthetic system, and therefore we consider, that the use of quantum simulations may be  a feasible alternative in systems with  {medium-strong coupling} and \emph {non-Markovian systems}, in the future. However, the algorithm obtained, due to the high requirements of gates and qubits, is not scalable to {real world} photosynthetic systems, with the current state of quantum technology. Hence, this simulation should be seen as a {proof of concept}, since a realistic quantum simulation of a photosynthetic system would have to involve hundreds of light-harvesting molecules, which is beyond the current quantum technology. Furthermore, the algorithm only effectively simulates pure-dephasing baths. For future work, we aim at extending it to new types of bath, e.g. those allowing for higher exciton recombination rates and non-Markovian effects as well as to new geometries of photosynthetic systems, in particular, to the Fenna-Matthews-Olson complex \cite{mohseni2014quantum}.

\section*{Competing Interests}

The authors declare that there are no conflicts of interest.

\section*{Acknowledgements}

Carlos Tavares was funded by the FCT -- Funda{\c{c}}{\~a}o para a Ci{\^e}ncia e Tecnologia (FCT) by the grant SFRH/BD/116367/2016, funded under the POCH programme and MCTES national funds.  This work is also financed by the ERDF – European Regional Development Fund through the Operational Programme for Competitiveness and Internationalisation - COMPETE 2020 Programme and by national funds through the Portuguese funding agency, FCT - Fundação para a Ciência e a Tecnologia, within project KLEE (POCI-01-0145-FEDER-030947) and the Strategic Funding UIDB/04650/2020 of the Centre of Physics. 

\bibliographystyle{plain}

\bibliography{references}
\newpage

\section*{Supplementary Information}

\subsection* {A.1 - Brief introduction to quantum systems}\label{Quantum systems}

There are several formalisms in which quantum mechanics can be expressed, including the one making use of Hilbert spaces and density matrices \cite {nielsen2002quantum},  which we shall denote as \emph {Hilbert} for clarity.
In the \emph {Hilbert} formalism, the theory is expressed in terms of states, $\ket {\Psi}$, and transitions between them.

\begin{figure}[H]
        \centering
        \includegraphics[width=0.3\linewidth]{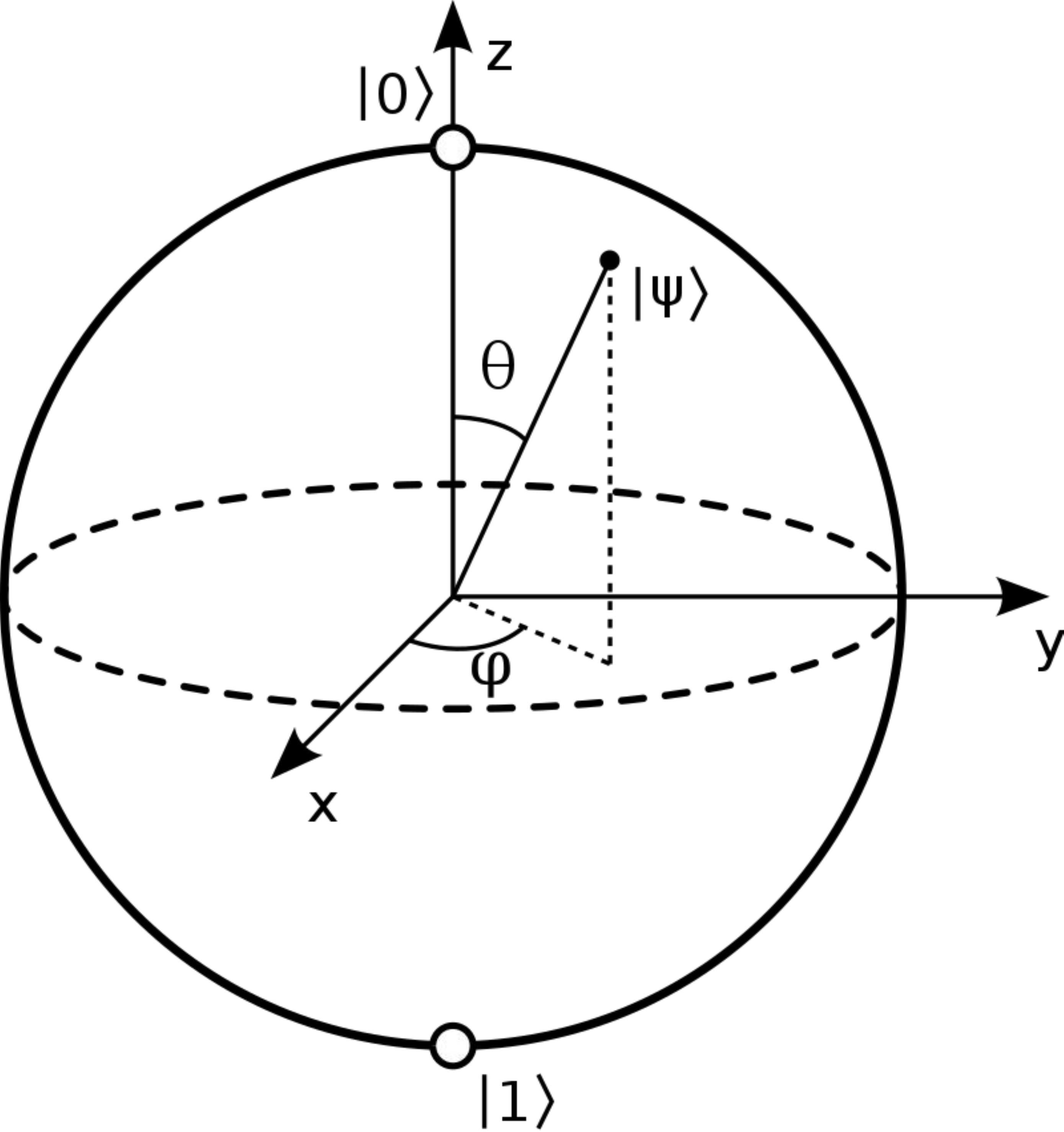}
        \caption{The set of points definable on the surface of the Bloch sphere, compose the state space of a \emph {Hilbert} space of dimension 2 (\emph {qubit}) of basis $\ket {0}$, $\ket {1}$, i.e. $\ket {\Psi} = \lambda \ket {0} + \beta \ket {1}$ with $|\lambda|^2 + |\beta|^2=1$.}
        \label {fig:bloch}
        
    \end{figure}

The states may be expanded on an \emph {eigenbasis} of the Hilbert space,

\begin {equation*}
\ket {\Psi} = \sum_i \lambda_i \ket {\Psi_i}, \text {with } \lambda_i \in \mathbb {C}
\end {equation*}

\noindent where $\sum_i |\lambda_i|^2 = 1$. The \emph {eigenbasis} states are often taken as the \emph {eigenstates} of a Hamiltonian and known in quantum mechanics as the \emph {stationary states}, while $\ket {\Psi}$ is an example of the so-called \emph {superposition states}. 

Furthermore, quantum systems can also be composed and form new systems through the \emph {Kronecker} product of the Hilbert spaces, corresponding to two sub-systems (in this case 1 and 2):

\begin {equation}
\mathcal{H}_1 \otimes \mathcal{H}_2
\end {equation}

\noindent The analysis of this operator quickly leads to another very relevant property of quantum mechanics, which is the existence of \emph {non-separable} states, i.e. states that cannot be written as a Cartesian separable product of the subsystems $1$ and $2$, as for instance happens in the so-called Bell states,
\begin {equation}
\ket {\Psi} = \frac {1}{\sqrt {2}} \left(\ket {0_10_2} + \ket {1_11_2} \right)\; ,
\end {equation}
\noindent which are a possible state of the $\otimes$ on two qubits. This phenomenon is the \emph {so-called} \emph {entanglement}, a vital component of many quantum technologies such as quantum information. An extension of \emph {pure} states represented by vectors in the Hilbert space leads to the density matrices, which can also describe the so called \emph {mixed} states. A mixed state can be considered as an mixture of several pure states, with certain statistical weights, therefore it cannot be described by a vector in the Hilbert space. A (Hermitian) density matrix (or operator) can be written as
\begin{equation*}
\hat \rho = \sum _{\alpha }P_\alpha\ket{\Phi_\alpha} \bra {\Phi_\alpha}
\end{equation*}
\noindent where $P_\alpha$ are real numbers between 0 and 1 and $\ket {\Phi _\alpha}\bra{\Phi _\alpha}$ is known as the \emph {projection} operator of a state $\Phi _\alpha$. For a \emph {pure} state all $\Psi _\alpha =0$ except for one, which is equal to unity. By expanding $\Phi _\alpha$ on a basis of stationary states, one can obtain the matrix representation of $\hat \rho$ in that basis.   
The density matrix of a system composed of two parts, $A$ and $B$, allows for a simple criterion for checking whether these parts are entangled or not in a particular state, $\hat \rho_{A+B}$ (see, e.g. \cite {barnett2009quantum}).
To distinguish between entangled and separable states, one needs to calculate the density matrix of subsystem $A$ (or $B$) as the partial trace of $\hat \rho_{A+B}$ and check the trace of $\hat \rho_{A}^2$. If it is equal to unity, the subsystem $A$ is in a pure state and the whole system is separable.    

%




While the Hamiltonian formalism of quantum mechanics permits properly describe \emph {closed} systems (and, in principle, universe as a whole), in practice, for instance in quantum technology, one deals with systems that are not closed, i.e. there is a free uncontrollable element, not necessarily conservative, interacting the system under study. The theory that studies these kind of systems is a branch of quantum mechanics known as \emph {open quantum systems} \cite {breuer2002theory}. In fact, an open system is nothing more than a \emph {closed system} composed of two subsystems, the \emph {system} and the \emph {environment}, in which the system plus environment Hilbert space is expressed as $\mathcal {H}_{S+E}$.

\noindent The overall effect of the environment in the system can be obtained by a \emph {tracing} operation over the environment Hilbert space:
\begin {equation}
\mathcal {H}_S = Tr_{E} [\mathcal{H}_{S+E}] \;. 
\end {equation}

\subsection*{A.2 - Solution of the Schr\"{o}dinger and Liouville equations for the two-molecule system}
Making use of the Schr\"{o}dinger equation ($\hbar=1$):
\begin{equation}
i\frac{\delta}{\delta t}\Ket{\Psi}=\hat{H}\Ket{\Psi}
\end{equation}
and writing the state $\Ket{\Psi}$ in the site basis as:
\begin{equation}
\Ket{\Psi}=\alpha \Ket{0} + \beta \Ket{1}
\end{equation}
where $|\alpha|^2+|\beta|^2=1$. One gets by applying the Schr\"{o}dinger equation to the state $\Ket{\Psi}$ and multiplying it by $\Bra{0}$ and $\Bra{1}$:
\begin{equation}
i\dot{\alpha} = \epsilon_{0}\alpha + J\beta \label{x1}
\end{equation}
\begin{equation}
i\dot{\beta} = \epsilon_{1}\beta + J\alpha  \label{x2}
\end{equation}
By making the substitions $\alpha=a e^{-i\epsilon_{0}t}$ and $\beta=b e^{-i\epsilon_{1}t}$ so that:
\begin{equation}
\dot{\alpha}=\dot{a}e^{-i\epsilon_{0}t}-i\epsilon_{0}\alpha
\end{equation}
\begin{equation}
\dot{\beta}=\dot{b}e^{-i\epsilon_{1}t}-i\epsilon_{1}\beta
\end{equation}
one has, taking into account the substitutions and Eqs. (\ref{x1}) and (\ref{x2}):
\begin{equation}
i\dot{a} = Je^{-iwt}b \label{x3}
\end{equation}
\begin{equation}
i\dot{b} = Je^{iwt}a \label{x4}
\end{equation}
where $w=\epsilon_{1}-\epsilon_{0}$. Then the variable $c$ is introduced as $c=ae^{iwt}$. One can observe through Eq. (\ref{x3}) that $\dot{a}=(\dot{c}-iwc)e^{-iwt}$ so that $i(\dot{c}-iwc)=Jb$. Differentiating with respect to time and using Eq. (\ref{x4}), one has:
\begin{equation}
\ddot{c}-iw\dot{c}+J^{2}c=0 \label{x5}
\end{equation}
By seeking solutions as $c=e^{i\lambda t}$:
\begin{equation}
-\lambda^{2}+w\lambda+J^{2}=0
\end{equation}
one gets:
\begin{equation}
\lambda_{0,1}=\frac{w}{2}\pm \sqrt{\frac{w^{2}}{4}+J^{2}}=\frac{1}{2}(w\pm \Omega)
\end{equation}
where $\Omega=\sqrt{w^{2}+4J^{2}}$. The solution of Eq. (\ref{x5}) is:
\begin{equation}
c=Ae^{i\lambda_{0}t}+Be^{i\lambda_{1}t}
\end{equation}
then,
\begin{equation}
a(t)=Ae^{i\frac{\Omega-w}{2}t}+Be^{-i\frac{\Omega+w}{2}t}
\end{equation}
and $b(t)$ is found by using equation (\ref{x4}):
\begin{equation}
b(t)=\frac{\Omega+w}{2J}Be^{i\frac{w-\Omega}{2}t}- \frac{\Omega-w}{2J}Ae^{i\frac{w+\Omega}{2}t}
\end{equation}
The constants $A$ and $B$ depend on the initial conditions. If the donor molecule is the only one excited ($\Ket{0}$) at the initial time $t=0$, then $\alpha(0)=1$ and $\beta(0)=0$. Thus, $a(0)=1$ and $b(0)=0$ and one gets $A=\frac{\Omega+w}{2\Omega}$ and $B=\frac{\Omega-w}{2\Omega}$. The solutions are:
\begin{equation}
a(t)=\frac{\lambda_{0}}{\Omega}e^{-i\lambda_{0}t}-\frac{\lambda_{1}}{\Omega}e^{-i\lambda_{1}t}
\end{equation}
\begin{equation}
b(t)=\frac{J}{\Omega}[e^{i\lambda_{0}t}-e^{i\lambda_{1}t}] \label{equation}
\end{equation}
and the system's density matrix becomes:
\begin{equation}
\rho=\Ket{\Psi}\Bra{\Psi}=\begin{pmatrix}
|\alpha|^2 & \alpha \beta^{\dag} \\
\alpha^{\dag} \beta & |\beta|^2
\end{pmatrix} = \begin{pmatrix}
|a(t)|^2 & a(t) b^{\dag}(t)e^{iwt} \\
a^{\dag}(t)b(t) e^{-iwt} & |b(t)|^2
\end{pmatrix}\;.
\end{equation}

The time dependence of the non-diagonal elements of the density matrix in the energy basis can be obtained much easier.
Consider the quantum master Liouville equation applied to a closed system $\rho$ subjected to a Hamiltonian $H$, defined as:
\begin{equation}
\frac{d\rho}{dt} = -i [H,\rho]
\end{equation}
Writing the density matrix in the energy eigenbasis (where the Hamiltonian is diagonal) yields:
\begin{equation}
\frac{d\rho_{ij}}{dt} = -i (E_{i}-E_{j})\rho_{ij}\;,
\end{equation}
where $E_{i}$ is the energy eigenvalue of the $i$th energy eigenstate. The populations (matrix elements $\rho_{ij}$, where $i=j$) remain constant and the coherence terms (matrix elements $\rho_{ij}$, where $i\neq j$) evolve oscillating in time as:
\begin{equation}
\rho_{ij}(t) = e^{-i (E_{i}-E_{j})t}\rho_{ij}(0) \label{coher}
\end{equation}
For the two molecule system, the Hamiltonian 

\begin{equation}
\hat{H}_{S}=\begin{pmatrix}
\epsilon_{0} & J \\
J & \epsilon_{1} 
\end{pmatrix} \label{diagH} \;.
\end{equation}

\noindent the energy eigenvalues $E_{0}$ and $E_{1}$ become, upon diagonalization, as
\begin{equation}
E_{0}=\frac{1}{2}[\epsilon_{0}+\epsilon_{1} + \sqrt{(\epsilon_{0}-\epsilon_{1})^{2}+4J^{2}}]
\end{equation}
and
\begin{equation}
E_{1}=\frac{1}{2}[\epsilon_{0}+\epsilon_{1}- \sqrt{(\epsilon_{0}-\epsilon_{1})^{2}+4J^{2}}]\;.
\end{equation}
Therefore, the coherence terms in Eq. \eqref{coher} evolve as:
\begin{equation}
\rho_{ij}(t) = e^{-i \omega t}\rho_{ij}(0)
\end{equation}
where the frequency of the quantum beating is given by $\Omega=\sqrt{(\epsilon_{0}-\epsilon_{1})^{2}+4J^{2}}$. 

\noindent This corresponds to the equation,
\begin{equation}
\rho_{ij}(t) = e^{-{it} \sqrt{(\epsilon_{0}-\epsilon_{1})^{2}+4J^{2}}/\hbar }\rho_{ij}(0)\,,\qquad i \neq j \,
\label{eq:rho}
\end{equation}
\noindent ,presented in the main text. The period of the oscillations is $\frac{2\pi}{\Omega}$.

\subsection*{A.3 - Circuits for the coherent regime} 

\begin{figure}[H]
	\centering
	\includegraphics[width=1\linewidth]{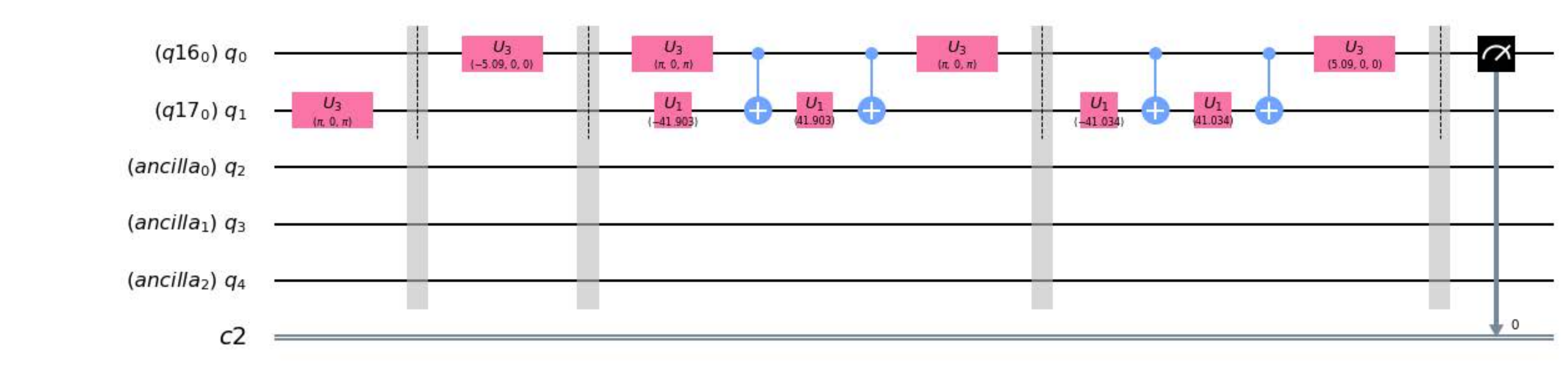}
	\caption{Optimized quantum circuit for the near-resonance system simulation.}
	\label{fig:optqcres}
\end{figure}

\begin{figure}[H]
	\centering
	\includegraphics[width=1\linewidth]{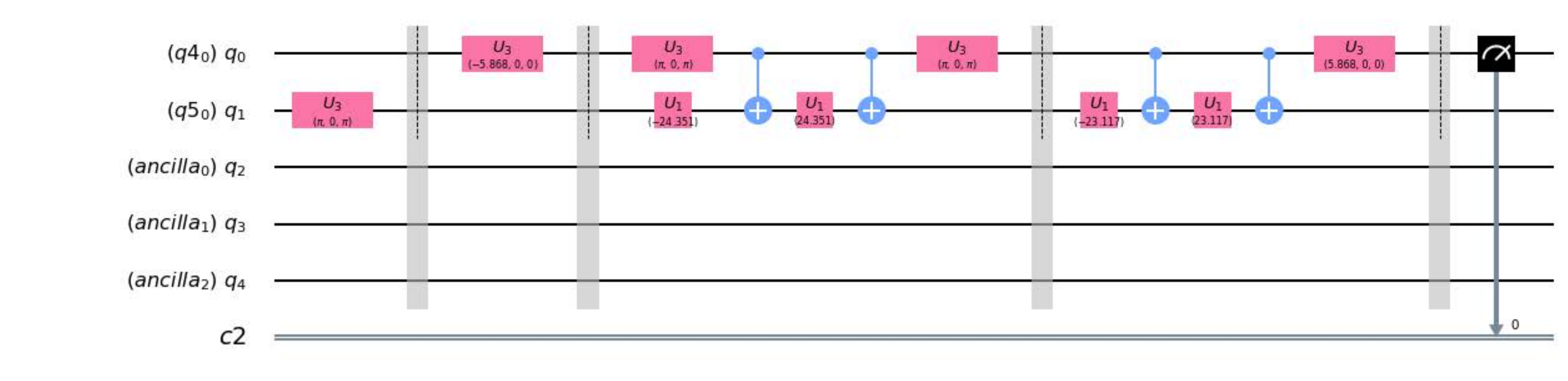}
	\caption{Optimized quantum circuit for the non-resonance system simulation.}
	\label{fig:optqcnres}
\end{figure}

This annex contains the optimized circuits  used to build the quantum simulations of section \emph{No decoherence regime} for the near resonance (figure \ref {fig:optqcres}) and off resonance (figure \ref {fig:optqcnres}) regimes.

\label {nonresson}

\end{document}